\documentclass[prd,aps,nofootinbib,tightenlines]{revtex4}
\usepackage{mathrsfs}
\usepackage{amsmath}
\usepackage{amssymb}
\usepackage{epsfig}
\usepackage{graphicx}
\usepackage{booktabs}
\usepackage{multirow}
\usepackage{subfigure}
\usepackage[section]{placeins}
\usepackage{float}
\usepackage{slashed}
\usepackage{longtable}
\usepackage{footnote}
\usepackage[colorlinks=true,linkcolor=red,citecolor=blue]{hyperref}
\begin{document}
\newcommand{\psl}{ p \hspace{-1.8truemm}/ }
\newcommand{\nsl}{ n \hspace{-2.2truemm}/ }
\newcommand{\vsl}{ v \hspace{-2.2truemm}/ }
\newcommand{\epsl}{\epsilon \hspace{-1.8truemm}/\,  }
\title{Cabibbo-suppressed charged-current  semileptonic decays of $\Xi_b$ baryons}
\author{Zhou Rui$^1$ }\email{jindui1127@126.com}
\author{Zhi-Tian Zou$^1$ }\email{zouzt@ytu.edu.cn}
\author{Ya Li$^2$}\email{liyakelly@163.com}
\author{Ying Li$^1$ }\email{liying@ytu.edu.cn}
\affiliation{Department of Physics, Yantai University, Yantai 264005, China}
\affiliation{Department of Physics and Institute of Theoretical Physics, Nanjing Normal University, Nanjing 210023, China}
\date{\today}
\begin{abstract}
We present the first perturbative QCD calculations of the $\Xi_b \to (\Lambda, \Sigma)$ transition form factors at leading order in $\alpha_s$, which govern the Cabibbo-suppressed semileptonic decays $\Xi_b \to (\Lambda, \Sigma)\ell \nu_\ell$ with $\ell = e, \mu, \tau$. Using these form factors, we evaluate differential and integrated branching fractions and angular observables within the helicity formalism. The branching ratios are predicted to be of order $10^{-4}$ for $\Sigma$ final states and $10^{-5}$ for $\Lambda$ final states, making them accessible to ongoing experiments such as LHCb. Ratios of decay rates between $\tau$ and $e$ channels are also provided, offering new probes of lepton-flavor universality. Lepton-mass effects are found to significantly impact the integrated angular observables. Furthermore, a combined analysis of $b \to u$ and $b \to c$ transitions in $\Xi_b$ decays yields subpercent precision for the ratios $\mathcal{R}_\ell(\Sigma/\Xi_c)$, enabling an independent determination of $|V_{ub}/V_{cb}|$ once the relevant decay-rate measurements become available.
 
\end{abstract}

\maketitle
\section{Introduction}
Charged-current semileptonic decays of $b$-hadrons are of great phenomenological significance as they provide an ideal place to constrain the Cabibbo-Kobayashi-Maskawa (CKM) matrix elements $V_{ub}$ or $V_{cb}$. Their measurements are essential for understanding the weak interactions in the Standard Model (SM) and probing potential new physics (NP). Currently, the determinations of $V_{ub}$ and $V_{cb}$ come mainly from the experimental measurements of the semileptonic decays of the $B$ meson at the $B$ factories, including the exclusive and inclusive processes~\cite{PDG2024,HFLAV2024}. However, there exists a tension between the determined values from the exclusive and inclusive semileptonic $B$ meson decays \cite{Kang:2013jaa, Belle:2016ure}, which leads to a long-standing inclusive versus exclusive puzzle in the heavy flavor sector of the SM~\cite{Bernlochner:2017xyx, Ricciardi:2019zph, Ricciardi:2014aya, Gambino:2019sif,Martinelli:2021onb,Martinelli:2021myh,Martinelli:2024bov,Martinelli:2023fwm}. Many efforts to ease the tension have been proposed both within and beyond the SM \cite{Buras:2010pz, Crivellin:2009sd, Crivellin:2014zpa, Bernlochner:2014ova, Bigi:2017njr, Altmannshofer:2021uub, Colangelo:2016ymy, Iguro:2020cpg, Bigi:2015uba, Buras:2013ooa, Bansal:2021oon}, but the situation remains unclear~\cite{Martinelli:2023fwm}. For the relevant reviews, see for example Refs.~\cite{Ricciardi:2021shl, Gambino:2020jvv} and references therein.

In addition to the aforementioned puzzle, the precise determination of the ratio of CKM matrix elements from measured decay-width ratios plays a crucial role in constraining the CKM unitarity triangle, since many experimental and theoretical uncertainties largely cancel in the ratio. Extractions of $|V_{ub}/V_{cb}|$ from combined studies of exclusive semileptonic $b\to u \ell\nu_\ell$ and $b\to c \ell\nu_\ell$ transitions in the $B$ sector have been discussed in~\cite{Bansal:2021oon, Biswas:2022yvh, LHCb:2020ist}. In the baryonic sector, the LHCb Collaboration has reported the first measurement of the ratio of the semileptonic $\Lambda_b\to p \mu \nu_\mu$ and $\Lambda_b\to \Lambda_c^+ \mu \nu_\mu$ decay rates in constrained kinematical regions~\cite{LHCb:2015eia}, providing a complementary determination of $|V_{ub}/V_{cb}|$ for the first time at a hadron collider. The measured $|V_{ub}|$ is consistent with world averages of exclusive measurements, but disagrees with the inclusive determination at the $3.5\sigma$ level. Thus, the LHCb result based on $\Lambda_b$ decays does not yet clarify the tension. On the theoretical side, combined analyses of $\Lambda_b\to p \ell \nu_\ell$ and $\Lambda_b\to \Lambda_c^+ \ell \nu_\ell$ have been performed within lattice QCD (LQCD)~\cite{Detmold:2015aaa}, the relativistic quark model (RQM)~\cite{Faustov:2016pal}, and various NP scenarios~\cite{Dutta:2015ueb}. The form factors obtained in these frameworks, when combined with LHCb data, enable determinations of $|V_{ub}|$ and $|V_{cb}|$. Interestingly, the LQCD-based extraction agrees well with exclusive $B$ decay results, whereas the RQM prediction aligns more closely with the inclusive determinations. A discrepancy exceeding $3\sigma$ remains between the two approaches. Although unresolved, these efforts are essential for advancing the determination of $|V_{ub}/V_{cb}|$. Continued experimental and theoretical progress may resolve this long-standing tension in the near future.

In principle, similar deviations from the SM predictions may also appear in other $b$ hadron decays. To gain a deeper understanding of the observed discrepancy, it is crucial to extract the CKM elements from a wider variety of decay modes. In particular, tree-level semileptonic decays of bottom baryons provide an independent determination of the relevant CKM matrix elements and can thus shed light on the origin of the discrepancy. As more data become available, an increasing number of $b$ baryon decay processes governed by the same underlying quark transitions can be systematically investigated.

$\Xi_b^0$ and $\Xi_b^-$ are the two antitriplet partners of $\Lambda_b$. In semileptonic $\Xi_b$ decays, the corresponding tree-level $b\to u(c) \ell \nu_\ell$ transitions are $\Xi_b^0\to \Sigma^+ (\Xi_c^+) \ell \nu_\ell$ and $\Xi_b^-\to \Lambda (\Xi_c^0) \ell \nu_\ell$. These processes serve as a valuable supplement to the well-studied $B$ meson and $\Lambda_b$ baryon decays, providing an alternative avenue for determining $|V_{ub}/V_{cb}|$, once the relevant hadronic form factors are established. Various theoretical approaches have been developed to calculate the form factors for $\Xi_b\to \Sigma, \Lambda, \Xi_c$ transitions~\cite{Cheng:1996cs, Ebert:2006rp, Singleton:1990ye, Cheng:1995fe, Ivanov:1996fj, Ivanov:1998ya, Cardarelli:1998tq, Albertus:2004wj, Korner:1994nh, Dutta:2018zqp, Zhang:2019xdm, Ke:2024aux, Neishabouri:2025abl, Azizi:2011mw}. The first attempt to extract $|V_{ub}/V_{cb}|$ by combining analyses of $\Xi_b^-\to \Lambda \ell \nu_\ell$ and $\Xi_b^-\to \Xi_c^0 \ell \nu_\ell$ was carried out in Ref.~\cite{Faustov:2018ahb}. In our previous work~\cite{Rui:2025iwa}, we investigated the semileptonic decays $\Xi_b\to \Xi_c \ell^- \bar{\nu}_\ell$ mediated by the $b\to c$ transition within the perturbative QCD (PQCD) framework, a well-established QCD-based approach for describing heavy-baryon decays~\cite{Lu:2009cm,Zhang:2022iun,Rui:2022sdc,Rui:2022jff,Rui:2023eac,Rui:2023fpp,Rui:2023fiz,Rui:2024xgc,Han:2022srw,Han:2024kgz,Han:2025tvc,Li:2025rsm,Yang:2025yaw}. A major advantage of the PQCD approach is the introduction of parton transverse momenta ($k_T$) to avoid end-point divergences, which allows for the calculation of many heavy-to-light form factors perturbatively, including their higher-twist corrections.
In principle,  from the perspective of effective field theory and QCD factorization, the hard function should only depend on the large scale, while soft scales like $k_T$ should be expanded  order by order in a strict power counting, with their effects systematically absorbed into higher-twist  light-cone distribution amplitudes (LCDAs). 
Recent efforts on the isolation of end-oint logarithms and  the refactorization applied to the end-point region have been achieved in Refs~\cite{Beneke:2022obx,Feldmann:2022ixt,Liu:2019oav,Liu:2020wbn,Lu:2022kos,Bell:2024bxg}. 
As pointed out in~\cite{Yu:2025jcs}, $k_T$ span three hierarchical scales such as hard, hard collinear, and soft, which should be handled at different characteristic scales separately.  It implies that the introduction of $k_T$ in the PQCD approach is equivalent to  a resummation of a certain class of contributions, which are supposed to be of higher twists. For a more in-depth discussion on this topic, one could refer to~\cite{Yu:2025jcs}. 
The current work focuses on the corresponding Cabibbo-suppressed decays mediated by the $b\to u$ transition. These channels are likewise governed by six form factors that parametrize the transition matrix elements of the low-energy effective Hamiltonian. In PQCD, these form factors can be evaluated through the convolution of the perturbative hard kernel with the  LCDAs of the initial and final states. Using the extracted form factors, we then calculate both differential and total decay rates for the considered channels. In addition, we examine the lepton-flavor universality (LFU) ratios between the $\tau$ and $e$ modes, which are potentially sensitive to NP effects~\cite{Li:2018lxi, Pich:2019pzg, Iguro:2024hyk}. Our results further allow a determination of $|V_{ub}/V_{cb}|$ once the relevant decay-rate ratios are experimentally measured. Beyond these, we also provide predictions for angular asymmetries, which may offer complementary probes of possible interactions beyond the SM.

The presentation of the paper is as follows. After this Introduction, we define the involved kinematic variables and present the transition form factors and helicity amplitudes in Sec.~\ref{sec:framework}. Section~\ref{sec:results} is devoted to the numerical analysis of the form factors as well as the semileptonic decay rates and other observables. We then briefly summarize in  Sec.~\ref{sec:sum}. The factorization formulas for the transition form factors are collected in the Appendix.

\section{ The baryonic form factors in PQCD}\label{sec:framework}
\begin{figure}[!htbh]
	\begin{center}
	   \vspace{0cm}  \centerline{\epsfxsize=25cm \epsffile{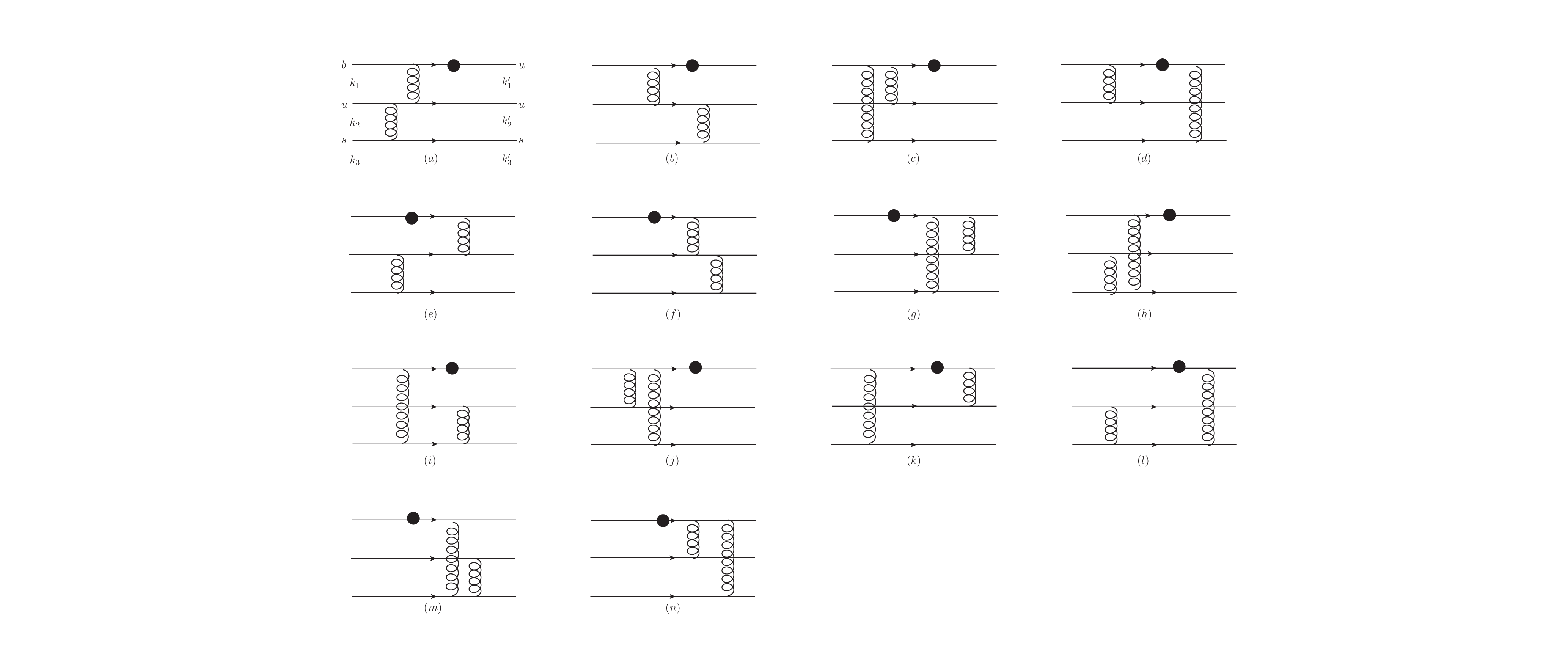}}
		\caption{
Lowest-order diagrams for $\Xi_b\to \Sigma$ transition form factors.
The symbol $\bullet$ denotes the electroweak vertex, from which the lepton pair emerges.}
		\label{fig:C}
	\end{center}
\end{figure}

This section is devoted to the calculation of form factors within the PQCD framework. As mentioned earlier, these channels proceed through the $b\to u$ transition at the quark level. The corresponding low-energy effective Hamiltonian governing this transition can be expressed as
\begin{eqnarray}
\mathcal{H}_{eff}=\frac{G_F}{\sqrt{2}}V_{ub}  \left[\bar u\gamma_\mu (1-\gamma_5)b \bar \ell \gamma^\mu (1-\gamma_5)\nu_\ell\right],
\end{eqnarray}
where $G_F$ is the Fermi constant, and $V_{ub}$ is the CKM matrix element. In order to get the amplitudes, we need to sandwich the effective Hamiltonians between the initial and final states. Then the corresponding amplitude can be factorized into the hadronic and leptonic contributions as
\begin{eqnarray}
\mathcal{M}=\frac{G_F}{\sqrt{2}}V_{ub} \langle \mathcal{B}_f(p')| \bar u\gamma_\mu (1-\gamma_5)b|\mathcal{B}_i(p)\rangle \bar \ell \gamma^\mu (1-\gamma_5)\nu_\ell.
\end{eqnarray}
Here we shall employ a generic notation such that the parent and daughter baryons are denoted by $\mathcal{B}_{i}$ and $\mathcal{B}_{f}$, respectively. $p$ and $p'$ are their respective momenta and $q=p-p'$ is the momentum transferred to the lepton-antineutrino pair. The hadronic matrix elements $\langle \mathcal{B}_f(p')| \bar u\gamma_\mu (1-\gamma_5)b|\mathcal{B}_i(p)\rangle$ are key inputs to the SM calculation and can be parametrized by form factors, which are functions of $q^2$ and depend on the strong interaction effects that bind the quarks inside the baryons. The form factors for the weak charged-current transition are defined in the standard way as follows:
\begin{eqnarray}\label{eq:FFs}
\langle \mathcal{B}_f|\bar {u}(1-\gamma_5)\gamma_\mu b |\mathcal{B}_i\rangle&=&\bar{u}_{\mathcal{B}_f}(p',\lambda') \left[\gamma_\mu f_1(q^2)- i\sigma_{\mu\nu}\frac{q^\nu}{M}f_2(q^2)+\frac{q_\mu}{M}f_3(q^2)\right]u_{\mathcal{B}_i}(p,\lambda)\nonumber\\
&+&\bar{u}_{\mathcal{B}_f}(p',\lambda') \left[\gamma_\mu g_1(q^2)- i\sigma_{\mu\nu}\frac{q^\nu}{M}g_2(q^2)+\frac{q_\mu}{M}g_3(q^2)\right]\gamma_5u_{\mathcal{B}_i}(p,\lambda),
\end{eqnarray}
where the labels $\lambda$ and $\lambda'$ denote the helicities of the two baryons. $u_{\mathcal{B}_{i(f)}}$ is the Dirac spinor of the initial (final) baryon. $f_i$ and $g_i$ with $i=1,2,3$ are three dimensionless vector and axial transitions form factors, respectively. The form factors $f_3$ and $g_3$ will contribute to  the tauonic modes. Using the equations of motion,  Eq.~(\ref{eq:FFs}) can be rewritten as
\begin{eqnarray}
\langle \mathcal{B}_f|\bar {u}(1-\gamma_5)\gamma_\mu b |\mathcal{B}_i\rangle&=&\bar{u}_{\mathcal{B}_f}(p',\lambda') \left[\gamma_\mu F_1(q^2)+F_2(q^2)\frac{p_\mu}{M}+F_3(q^2)\frac{p'_\mu}{M}\right]u_{\mathcal{B}_i}(p,\lambda),\nonumber\\
&+&\bar{u}_{\mathcal{B}_f}(p',\lambda') \left[\gamma_\mu G_1(q^2)+G_2(q^2)\frac{p_\mu}{M}+G_3(q^2)\frac{p'_\mu}{M}\right]\gamma_5u_{\mathcal{B}_i}(p,\lambda),
\end{eqnarray}
with
\begin{eqnarray}
f_1&=&F_1+\frac{1}{2}(F_2+F_3)(1+r), \quad f_2=-\frac{1}{2}(F_2+F_3), \quad f_3=\frac{1}{2}(F_2-F_3),\nonumber\\
g_1&=&G_1-\frac{1}{2}(G_2+G_3)(1-r), \quad g_2=-\frac{1}{2}(G_2+G_3), \quad g_3=\frac{1}{2}(G_2-G_3).
\end{eqnarray}
Here, $r=m/M$ is the mass ratio between the final and initial baryons. The details of how to calculate $F_i$ and $G_i$ for a heavy-to-heavy transition in PQCD were discussed in our previous paper~\cite{Rui:2025iwa}. This formalism can be extended to the heavy-to-light decays straightforwardly, whose factorization formula can be written as
\begin{eqnarray}\label{eq:FG}
F_i/G_i=\frac{4f_{\mathcal{B}_i}  \pi^2 G_F}{27\sqrt{2}}\sum_\xi
\int\mathcal{D}x\mathcal{D}b
\alpha_s^2(t_\xi)e^{-S_{\mathcal{B}_i}-S_{\mathcal{B}_f}}\Omega_\xi(b_i,b'_i) H^{F_i/G_i}_\xi(x_i,x'_i),
\end{eqnarray}
with the integration measures  
\begin{eqnarray}
\mathcal{D}x&=&dx_1dx_2dx_3\delta(1-x_1-x_2-x_3)dx'_1dx'_2dx'_3\delta(1-x'_1-x'_2-x'_3),\nonumber\\
\mathcal{D}b&=& d^2\textbf{b}_2d^2\textbf{b}_3d^2\textbf{b}'_2d^2\textbf{b}'_3.
\end{eqnarray}
$x^{(\prime)}_i$ with $i=1,2,3$ represents the longitudinal momentum fractions of the valence quarks inside the baryons, and $\textbf{b}^{{(\prime)}}_{i}$ are the conjugate variables for the corresponding transverse momentum $\textbf{k}^{{(\prime)}}_{iT}$ in the impact $b$ space. The summation extends over all possible diagrams as shown in Fig.~\ref{fig:C}. Note that Figs.~\ref{fig:C}(h)-\ref{fig:C}(n) are new compared to the $\Xi_b\to \Xi_c$ process in~\cite{Rui:2025iwa} because the symmetry of the baryonic wave function  is not applicable for the topologies in Fig.~\ref{fig:C}. We shall demonstrate that these new diagrams, in fact, give different contributions than those from Fig.~\ref{fig:C}(a)-\ref{fig:C}(g).
The Sudakov exponent $S_{\mathcal{B}_{i(f)}}$ in Eq.~(\ref{eq:FG}) arises from the resummation $k_T$ of large logarithmic corrections to baryonic LCDAs, which suppresses long-distance contributions from the small $k_T$ region, and improves the applicability of PQCD to decays of $b$ hadrons. Their explicit forms can be found in Refs.~\cite{Rui:2025iwa,Lu:2009cm}. $\Omega_{\xi}$ is the Fourier transformation of the denominator of the hard amplitude from the $k_T$ space to its conjugate $b$ space. $H_{\xi}$ is the numerator of the hard amplitude depending on the spin structure of final state. These quantities together with the  hard scales $t_\xi$ are provided in Appendix~\ref{sec:for}.

The calculations of $H_{\xi}$  require two nonperturbative inputs such as the LCDAs of the initial and final baryons, which can be constructed via the nonlocal matrix elements. The LCDAs of the $\Xi_b$ baryon in the heavy quark limit can be constructed similar to the $\Lambda_b$ one~\cite{plb665197, jhep112013191, epjc732302, plb738334, jhep022016179, Ali:2012zza,Han:2024min,Han:2024yun,Wang:2024wwa}. Four popular models of $\Xi_b$ baryon LCDAs up to twist four~\cite{jhep112013191,plb665197,epjc732302}  have been collected in~\cite{Rui:2025iwa}. The explicit expressions for  the LCDAs of $\Sigma$ and $\Lambda$ baryons up to twist six are obtained in~\cite{Liu:2008yg}, and, for completeness, they are presented in Appendix~\ref{sec:LCDAs}.

The calculations are performed in the rest frame of the parent baryon $\mathcal{B}_i$ with the daughter baryon $\mathcal{B}_f$ moving in the plus direction. In the light-cone coordinates, the momentum $p$ and $p'$ can be denoted as 
\begin{eqnarray}
p=\frac{M}{\sqrt{2}}\left(1,1,\textbf{0}_{T}\right), \quad p'=\frac{M}{\sqrt{2}}\left(f^+,f^-,\textbf{0}_{T}\right),
\end{eqnarray}
with the factors $f^{\pm}=(f\pm \sqrt{f^2-1})r$ and $f=\frac{p \cdot  p'}{Mm}$. The momenta of the valence quarks in the $\Xi_b$ baryon are
\begin{eqnarray}
k_{2,3}=\left(0,\frac{M}{\sqrt{2}}x_{2,3},\textbf{k}_{iT}\right), \quad k_1=p-k_2-k_3,
\end{eqnarray}
where $k_1$ is associated with the $b$ quark. The momenta of the valence quarks in the daughter baryon are
\begin{eqnarray}
k'_i=\left(\frac{M}{\sqrt{2}}f^+x'_i,0,\textbf{k}'_{iT}\right).
\end{eqnarray}
Here, only $b$ quark is considered to be massive, while other light quarks are treated as massless. This means that, except for $k_1$, only one of the dominant components are kept for all other light quarks momenta in the massless  limit~\cite{Rui:2022sdc}. As the fast recoiled daughter baryon  moves approximately in the plus direction, $f^-\sim \mathcal{O}(r^2) $ is a small quantity and the sum of $k'_i$  equal to $p'$ holds approximately. Collecting all the above ingredients, we obtain the form factors in Eq.~(\ref{eq:FG}) in PQCD formulism.

In the helicity representation, a helicity amplitude including the vector ($V$) and the axial-vector ($A$) currents can be written as
\begin{eqnarray}
H_{\lambda'\lambda_W}=H^{V}_{\lambda'\lambda_W}-H^{A}_{\lambda'\lambda_W}=\langle \mathcal{B}_f(p')| \bar u\gamma_\mu (1-\gamma_5) b|\mathcal{B}_i(p)\rangle  \epsilon^{\ast \mu}(\lambda_W),
\end{eqnarray}
where $\epsilon^{\mu}(\lambda_W)$ is the polarization vector of virtual $W$ boson with $\lambda_W$ being its helicity component. It has four helicities, with two angular momentum $J=1,0$ in the rest frame of $W$, namely, $\lambda_W=0,\pm1 (J=1)$ and $\lambda_W=0 (J=0)$. The spin-0 component does not have effects on semileptonic decay in the lepton massless limit. To distinguish the two $\lambda_W=0$ states, we introduce $\lambda_W=t$ for $J=0$ with $t$ meaning temporal. If we choose  the daughter baryon pointing in the $+z$ direction and  $W$ pointing in the $-z$  direction, the helicity of the initial baryon, $\lambda$, is established by the relation $\lambda=\lambda'-\lambda_W$ according to the angular momentum conservation. Considering the restriction $\lambda=\pm \frac{1}{2}$, one has four independent vector helicity amplitudes
\begin{eqnarray}
H^V_{\frac{1}{2}0}&=&\sqrt{\frac{Q_-^2-q^2}{q^2}}[Q_+f_1(q^2)+\frac{q^2}{M}f_2(q^2)],\nonumber\\
H^A_{\frac{1}{2}0}&=&\sqrt{\frac{Q_+^2-q^2}{q^2}}[Q_-g_1(q^2)-\frac{q^2}{M}g_2(q^2)],\nonumber\\
H^V_{\frac{1}{2}1}&=&\sqrt{2(Q_-^2-q^2)}[-f_1(q^2)-\frac{Q_+}{M}f_2(q^2)],\nonumber\\
H^A_{\frac{1}{2}1}&=&\sqrt{2(Q_+^2-q^2)}[-g_1(q^2)+\frac{Q_-}{M}g_2(q^2)],
\end{eqnarray}
and two scalar helicity amplitudes
\begin{eqnarray}
H^V_{\frac{1}{2}t}&=&\sqrt{\frac{Q_+^2-q^2}{q^2}}[Q_-f_1(q^2)+\frac{q^2}{M}f_3(q^2)],\nonumber\\
H^A_{\frac{1}{2}t}&=&\sqrt{\frac{Q_-^2-q^2}{q^2}}[Q_+g_1(q^2)-\frac{q^2}{M}g_3(q^2)],
\end{eqnarray}
where we abbreviate $Q_{\pm}=M\pm m$.  The helicity-flipped amplitudes can be obtained from the parity consideration or from explicit calculation, i.e., $H^V_{-\lambda'-\lambda_W}=H^V_{\lambda'\lambda_W}$ and $H^A_{-\lambda'-\lambda_W}=-H^A_{\lambda'\lambda_W}$~\cite{Faustov:2016pal}.  The total left-chiral helicity amplitude can be written as $H_{\lambda'\lambda_W}=H^V_{\lambda'\lambda_W}-H^A_{\lambda'\lambda_W}$.

The differential decay width for the semileptonic decay $\mathcal{B}_i\to \mathcal{B}_f \ell \nu_\ell$ is given by
\begin{eqnarray}\label{eq:gam}
\frac{d\Gamma_\ell}{dq^2}&=&\frac{G_F^2|V_{ub}|^2q^2|P|}{192M^2\pi^3}(1-2\delta_\ell)^2H,
\end{eqnarray}
where we have introduced the helicity-flip penalty factor $\delta_\ell=\frac{m_\ell^2}{2q^2}$ with $m_\ell$ being the lepton mass. $|P|$ is the momentum of the outgoing baryon. $H$ denotes the total helicity amplitude squared, which is defined as
\begin{eqnarray}
H=(|H_{\frac{1}{2}1}|^2+|H_{-\frac{1}{2}-1}|^2+|H_{\frac{1}{2}0}|^2+|H_{-\frac{1}{2}0}|^2)(1+\delta_\ell)+3\delta_\ell(|H_{\frac{1}{2}t}|^2+|H_{-\frac{1}{2}t}|^2).
\end{eqnarray}
Note that the timelike leptonic polarization state requires a nonvanishing lepton mass. Hence, the scalar amplitudes $H_{\frac{1}{2}t}$ and $H_{-\frac{1}{2}t}$ always enter observables with  $m_\ell$ such that all physical observables are well defined in the limit  $m_\ell\to0$. The expressions of the leptonic forward-backward asymmetries ($A^\ell_{FB}(q^2)$), the convexity parameter ($C^\ell_F(q^2)$), the final hadron polarizations ($H^\ell(q^2)$), and the lepton polarizations ($P^\ell(q^2)$) for the $\tau$ and $e$ modes  are written as
\begin{eqnarray}\label{eq:AFB}
A^\ell_{FB}(q^2)&=&-\frac{3}{4H}(H_{TP}+4\delta_\ell H_{SL}), \nonumber\\
C^\ell_{F}(q^2) &=&\frac{3}{4}(1-2\delta_\ell)\frac{H_T-2H_L}{H},\nonumber\\
H^{\ell}_z(q^2)&=& \frac{H_{TP}+H_{LP}+\delta_\ell(H_{TP}+H_{LP}+3H_{SP})}{H}, \nonumber\\
H^{\ell}_x(q^2)&=& -\frac{3\pi}{4\sqrt{2}} \frac{H_{LT}-2\delta_\ell H_{STP}}{H},\nonumber\\
P^{\ell}_z(q^2)&=& -\frac{H_{T}+H_{L}-\delta_\ell(H_{T}+H_{L}+3H_{S})}{H}, \nonumber\\
P^{\ell}_x(q^2)&=& -\frac{3\pi}{4\sqrt{2}}\sqrt{\delta_\ell}\frac{H_{TP}-2H_{SL}}{H},
\end{eqnarray}
where  various parity conserving and violating helicity structures entering the above equations are defined as
\begin{eqnarray}\label{eq:H1}
H_T(q^2)&=&|H_{\frac{1}{2}1}|^2+|H_{-\frac{1}{2}-1}|^2,\nonumber\\
H_L(q^2)&=&|H_{\frac{1}{2}0}|^2+|H_{-\frac{1}{2}0}|^2,\nonumber\\
H_S(q^2)&=&|H_{\frac{1}{2}t}|^2+|H_{-\frac{1}{2}t}|^2,\nonumber\\
H_{SL}(q^2)&=&Re[H_{\frac{1}{2}0}H^{\dag}_{\frac{1}{2}t}+H_{-\frac{1}{2}0}H^{\dag}_{-\frac{1}{2}t}],\nonumber\\
H_{ST}(q^2)&=&Re[H_{\frac{1}{2}1}H^{\dag}_{-\frac{1}{2}t}+H_{-\frac{1}{2}-1}H^{\dag}_{\frac{1}{2}t}],\nonumber\\
H_{LT}(q^2)&=&Re[H_{\frac{1}{2}1}H^{\dag}_{-\frac{1}{2}0}+H_{-\frac{1}{2}-1}H^{\dag}_{\frac{1}{2}0}],\nonumber\\
H_{TP}(q^2)&=&|H_{\frac{1}{2}1}|^2-|H_{-\frac{1}{2}-1}|^2,\nonumber\\
H_{LP}(q^2)&=&|H_{\frac{1}{2}0}|^2-|H_{-\frac{1}{2}0}|^2,\nonumber\\
H_{SP}(q^2)&=&|H_{\frac{1}{2}t}|^2-|H_{-\frac{1}{2}t}|^2,\nonumber\\
H_{SLP}(q^2)&=&Re[H_{\frac{1}{2}0}H^{\dag}_{\frac{1}{2}t}-H_{-\frac{1}{2}0}H^{\dag}_{-\frac{1}{2}t}],\nonumber\\
H_{STP}(q^2)&=&Re[H_{\frac{1}{2}1}H^{\dag}_{-\frac{1}{2}t}-H_{-\frac{1}{2}-1}H^{\dag}_{\frac{1}{2}t}],\nonumber\\
H_{LTP}(q^2)&=&Re[H_{\frac{1}{2}1}H^{\dag}_{-\frac{1}{2}0}-H_{-\frac{1}{2}-1}H^{\dag}_{\frac{1}{2}0}].
\end{eqnarray}
Note that when calculating these integrated observables, one has to remember to include a $q^2$-dependent kinematic factor $q^2|P|(1-\frac{m_l^2}{q^2})^2$ both in the numerator and denominator of Eq.~(\ref{eq:AFB}) and integrate over $q^2$ separately before taking the ratio. Their numerical results will be given in the next section.

\section{Numerical results}\label{sec:results}
\begin{table}[!htbh]
	\caption{The predicted form factors $f_i(q^2)$ and $g_i(q^2)$  at $q^2=0$ with four different $\Xi_b$ baryonic LCDAs for the $\Xi_b^0\to \Sigma^+$ transition. The fitted parameters are presented with only central values. The available predictions  in the previous literature are also shown for comparisons.}
\label{tab:form1}
	\begin{tabular}[t]{lccccc}
	\hline\hline
Form factor      & This work (S1)   & This work (S2)   &This work (S3)  &This work (S4) & ~\cite{Zhao:2018zcb}\footnotemark[1]  \\ \hline
$f_1(q^2=0) $    & $0.200^{+0.107+0.010+0.029}_{-0.053-0.009-0.023}$
                 & $0.208^{+0.013+0.012+0.027}_{-0.004-0.008-0.023}$
                 & $0.190^{+0.003+0.007+0.016}_{-0.002-0.010-0.017}$
                 & $0.263^{+0.147+0.008+0.031}_{-0.072-0.014-0.034}$                   &0.184   \\
$a_0 $           &0.361    &0.382  &0.369  &0.483 &$\cdots$    \\
$a_1 $           &$-1.121$   &$-1.229$ &$-1.269$ &$-1.566$ &$\cdots$    \\
$f_2(q^2=0) $    & $0.052^{+0.024+0.002+0.007}_{-0.012-0.002-0.005}$
                 &$0.053^{+0.004+0.003+0.007}_{-0.001-0.002-0.005}$
                  &$0.043^{+0.003+0.003+0.004}_{-0.003-0.002-0.004}$
                  &$0.067^{+0.033+0.004+0.008}_{-0.015-0.002-0.007}$                    &$-0.061$   \\
$a_0 $           &0.138    &0.144  &0.121  &0.179 &$\cdots$    \\
$a_1 $           &$-0.612$   &$-0.648$ &$-0.555$ &$-0.799$ &$\cdots$    \\
$f_3(q^2=0) $    & -$0.038^{+0.011+0.001+0.005}_{-0.018-0.002-0.007}$
                 & -$0.039^{+0.001+0.002+0.006}_{-0.002-0.002-0.006}$
                 & -$0.032^{+0.002+0.002+0.005}_{-0.003-0.002-0.004}$
                 & -$0.049^{+0.015+0.002+0.007}_{-0.030-0.002-0.008}$                    &$\cdots$    \\
$a_0 $           &$-0.109$   &$-0.116$  &$-0.101$ &$-0.143$ &$\cdots$    \\
$a_1 $           &0.505    &0.548   &0.489  &0.669 &$\cdots$    \\
$g_1(q^2=0) $    & $0.213^{+0.108+0.007+0.026}_{-0.057-0.009-0.027}$
                 & $0.217^{+0.015+0.010+0.025}_{-0.004-0.010-0.022}$
                 & $0.195^{+0.007+0.006+0.019}_{-0.000-0.011-0.015}$
                 & $0.274^{+0.148+0.011+0.036}_{-0.070-0.010-0.029}$                   &0.178   \\
$a_0 $           &0.407    &0.428  &0.427  &0.520 &$\cdots$    \\
$a_1 $           &$-1.386$   &$-1.493$ &$-1.659$ &$-1.746$ &$\cdots$    \\
$g_2(q^2=0) $    & $0.046^{+0.021+0.002+0.006}_{-0.011-0.002-0.006}$
                 & $0.048^{+0.004+0.002+0.006}_{-0.001-0.002-0.006}$
                  &$0.037^{+0.003+0.002+0.005}_{-0.002-0.002-0.005}$
                  &$0.059^{+0.029+0.002+0.007}_{-0.015-0.003-0.008}$                   &$-0.008$   \\
$a_0 $           &0.129    &0.135  &0.106  &0.169 &$\cdots$    \\
$a_1 $           &$-0.590$   &$-0.617$ &$-0.490$ &$-0.779$ &$\cdots$    \\
$g_3(q^2=0) $    & -$0.043^{+0.012+0.002+0.006}_{-0.023-0.002-0.006}$
                 & -$0.044^{+0.002+0.002+0.010}_{-0.003-0.002-0.005}$
                 & -$0.038^{+0.001+0.002+0.005}_{-0.003-0.002-0.005}$
                 & -$0.056^{+0.015+0.001+0.006}_{-0.033-0.003-0.008}$                  &$\cdots$     \\
$a_0 $           &$-0.117$   &$-0.129$  &$-0.112$ &$-0.156$ &$\cdots$    \\
$a_1 $           &0.519    &0.603   &0.525  &0.705 &$\cdots$    \\
		\hline\hline
	\end{tabular}
\footnotetext[1]{
The physical transition form factors are obtained  by multiplying the original values  by the corresponding overlap factor. }
\end{table}

This section is devoted to the numerical analysis of the form factors and various observables derived in the previous section. First let us specify the input parameters that are needed in performing the numerical calculations. The masses of the related baryons and leptons are $M_{\Xi_b}=5.797$ GeV, $m_{\Sigma}=1.189$ GeV, $m_{\Lambda}=1.116$ GeV, and $m_\tau=1.777$ GeV~\cite{PDG2024}. For the heavy $b$ quark mass, we use the $\overline{\text{MS}}$ scheme value, $m_b(\bar{m}_b)=4.18$ GeV~\cite{PDG2024}. The lifetimes of the baryons are taken as $\tau_{\Xi_b^-}=1.572$ ps and $\tau_{\Xi_b^0}=1.48$ ps~\cite{PDG2024}. 
For the   CKM matrix element $|V_{ub}|$, we use the value of $|V_{ub}|=4.05\times10^{-3}$~\footnote{ This value is somewhat higher than the Particle Data Group's recommended values of $V_{ub}=3.70\times 10^{-3}$ (exclusive) or $V_{ub}=3.82\times 10^{-3}$ (average), increasing the calculated branching ratios by about $10\% - 20\%$. } from Ref.~\cite{Faustov:2018ahb}  in order to eliminate the difference caused by the $V_{ub}$ when comparing our branching ratios with theirs. The input parameters entering the LCDAs of the $\Xi_b$ baryon with various models are collected in~\cite{Rui:2025iwa}, whose values are: 
\begin{eqnarray}\label{eq:mod}
 \left\{
            \begin{array}{ll}
               \omega_0=(0.4\pm0.1) \text{GeV} & \text{for Exponential model }, \\
\tau=(0.6\pm 0.2)      \text{GeV} & \text{for QCDSR model },\\
A=0.5\pm 0.2                       & \text{for Gegenbauer model },\\
\bar{\Lambda}=(0.8\pm 0.2)      \text{GeV} & \text{for Free parton model }.\\
            \end{array}
          \right.
\end{eqnarray}
The LCDAs of $\Sigma$ and $\Lambda$ baryons  each include four independent parameters, namely, $f_{\Lambda/\Sigma}$, $\lambda_1$, $\lambda_2$, and $\lambda_3$. These parameters have been calculated in the framework of the light-cone QCD sum rules~\cite{Liu:2008yg}, which are given as (GeV$^2$)
\begin{eqnarray}\label{eq:sigma}
f_{\Sigma}=(9.4\pm 0.4)\times 10^{-3}, \quad \lambda_1=-(2.5\pm0.1)\times 10^{-2}, \quad \lambda_2=(4.4\pm0.1)\times 10^{-2}, \quad \lambda_3=(2.0\pm0.1)\times 10^{-2},
\end{eqnarray}
for the $\Sigma$ baryon, and
\begin{eqnarray}\label{eq:lambda}
f_{\Lambda}=(6.0\pm 0.3)\times 10^{-3}, \quad \lambda_1=(1.0\pm0.3)\times 10^{-2}, \quad |\lambda_2|=(0.83\pm0.05)\times 10^{-2}, \quad |\lambda_3|=(0.83\pm0.05)\times 10^{-2},
\end{eqnarray}
for the $\Lambda$ baryon.

It should be emphasized that, PQCD are suitable for describing the low squared momentum transfer  $q^2$ region of the form factors. To broaden this restricted domain to the full physical domain given above, we use the $z$-series parametrization suggested in~\cite{Bourrely:2008za}, where
\begin{eqnarray}
z(q^2)=\frac{\sqrt{Q^2_+-q^2}-\sqrt{Q^2_+-Q^2_-}}{\sqrt{Q^2_+-q^2}+\sqrt{Q^2_+-Q^2_-}}.
\end{eqnarray}
The best parametrization of the form factors is to retain the first-order term,  
\begin{eqnarray}\label{eq:Fq}
f(q^2)/g(q^2)= \frac{a_0+a_1z(q^2)}{1-q^2/M^2_{\text{pole}}}.
\end{eqnarray}
where $a_{0}$ and $a_{1}$ are two free parameters  needed to be fitted. For the pole masses, we use~\cite{Faustov:2018ahb}
\begin{eqnarray}
M_{\text{pole}}= \left\{
            \begin{array}{ll}
               5.325~\text{GeV}, & \text{for}~ f_{1,2} \\
               5.723~\text{GeV}, & \text{for}~ g_{1,2} \\
               5.749~\text{GeV}, & \text{for}~ f_{3} \\
               5.280~\text{GeV}, & \text{for}~ g_{3} .\\
            \end{array}
          \right.
\end{eqnarray}

 In the fitting procedure, we first numerically evaluate each form factor at ten different $q^2$ points ranging from 0 to $m^2_\tau$ with PQCD, which are considered as a ten-point dataset. 
 The NonlinearModelFit function in Mathematica is used to determine the best-fit parameters $a_0$ and $a_1$ for the $z$-series model to match the dataset. It works by minimizing the difference between the model and the data. 

 As mentioned before, there are four available models for the $\Xi_b$ baryonic LCDAs,  namely Exponential~\cite{jhep112013191}, QCDSR~\cite{plb665197}, Gegenbauer~\cite{epjc732302}, and Free parton models~\cite{jhep112013191}.
 In the following numerical calculations, we mark the results calculated by using the four models as S1 (Exponential model), S2 (QCDSR model), S3 (Gegenbauer model), and S4 (Free parton model) schemes, respectively.
\begin{table}[!htbh]
	\caption{Same as Table~\ref{tab:form1} but for $\Xi_b^-\to \Lambda$.}
	\label{tab:form2}
	\begin{tabular}[t]{lcccccc}
	\hline\hline
Form factor      & This work (S1)   & This work (S2)   &This work (S3)  &This work (S4) & ~\cite{Zhao:2018zcb}\footnotemark[1] &~\cite{Faustov:2018ahb,Zhang:2019xdm} \\ \hline
$f_1(q^2=0) $    &$0.078^{+0.038+0.015+0.010}_{-0.019-0.015-0.010}$
                 &$0.078^{+0.004+0.016+0.009}_{-0.002-0.016-0.010}$
                 &$0.107^{+0.002+0.011+0.007}_{-0.000-0.011-0.007}$
                 &$0.100^{+0.048+0.018+0.010}_{-0.025-0.018-0.013}$                   &$-0.075$                &0.092\\
$a_0 $           &0.093    &0.094  &0.108  &0.131 &$\cdots$   &$\cdots$   \\
$a_1 $           &$-0.103$   &$-0.102$ &$-0.013$ &$-0.207$ &$\cdots$   &$\cdots$  \\
$f_2(q^2=0) $    &$0.028^{+0.007+0.005+0.003}_{-0.004-0.004-0.002}$
                 &$0.029^{+0.001+0.004+0.002}_{-0.001-0.005-0.002}$
                 &$0.036^{+0.001+0.003+0.002}_{-0.000-0.003-0.001}$
                 &$0.036^{+0.009+0.006+0.003}_{-0.005-0.006-0.003}$                   &0.025                 &0.029\\
$a_0 $           &0.066    &0.069  &0.078  &0.089 &$\cdots$   &$\cdots$   \\
$a_1 $           &$-0.255$   &$-0.264$ &$-0.277$ &$-0.349$ &$\cdots$   &$\cdots$  \\
$f_3(q^2=0) $    &$-0.020^{+0.004+0.004+0.002}_{-0.008-0.004-0.002}$
                 &$-0.020^{+0.000+0.004+0.002}_{-0.001-0.004-0.002}$
                 &$-0.027^{+0.000+0.003+0.001}_{-0.001-0.003-0.002}$
                 &$-0.025^{+0.005+0.006+0.003}_{-0.011-0.006-0.003}$                   &$\cdots$              &$-0.002$\\
$a_0 $           &$-0.049$   &$-0.053$ &$-0.062$ &$-0.063$ &$\cdots$   &$\cdots$  \\
$a_1 $           &0.191    &0.221  &0.229  &0.251  &$\cdots$   &$\cdots$   \\
$g_1(q^2=0) $    &$0.086^{+0.036+0.015+0.010}_{-0.019-0.016-0.010}$
                 &$0.087^{+0.004+0.017+0.010}_{-0.002-0.016-0.009}$
                 &$0.116^{+0.002+0.012+0.007}_{-0.000-0.011-0.006}$
                 &$0.113^{+0.049+0.021+0.012}_{-0.024-0.021-0.012}$                   &$-0.072$                &0.077\\
$a_0 $           &0.147    &0.142  &0.156  &0.175 &$\cdots$   &$\cdots$   \\
$a_1 $           &$-0.399$   &$-0.356$ &$-0.256$ &$-0.417$ &$\cdots$   &$\cdots$  \\
$g_2(q^2=0) $    &$0.029^{+0.007+0.003+0.002}_{-0.004-0.004-0.003}$
                 &$0.030^{+0.000+0.005+0.002}_{-0.000-0.005-0.002}$
                 &$0.038^{+0.001+0.004+0.002}_{-0.000-0.003-0.001}$
                 &$0.038^{+0.010+0.006+0.003}_{-0.004-0.005-0.003}$                   &0.003                 &0.007\\
$a_0 $           &0.079    &0.083  &0.086  &0.100  &$\cdots$   &$\cdots$   \\
$a_1 $           &$-0.329$   &$-0.346$ &$-0.312$ &$-0.416$ &$\cdots$   &$\cdots$  \\
$g_3(q^2=0) $    &$-0.018^{+0.004+0.004+0.002}_{-0.008-0.005-0.002}$
                 &$-0.018^{+0.000+0.003+0.001}_{-0.002-0.004-0.003}$
                 &$-0.025^{+0.001+0.003+0.003}_{-0.000-0.002-0.001}$
                 &$-0.023^{+0.005+0.005+0.003}_{-0.011-0.005-0.003}$                   &$\cdots$              &$-0.041$\\
$a_0 $           &$-0.038$   &$-0.043$ &$-0.040$ &$-0.052$ &$\cdots$   &$\cdots$  \\
$a_1 $           &0.131    &0.158  &0.102  &0.191  &$\cdots$   &$\cdots$   \\
		\hline\hline
	\end{tabular}
\footnotetext[1]{The physical transition form factors are obtained  by multiplying the original values  by the corresponding overlap factor. }
\end{table}
\begin{figure}[!htbh]
\begin{center}
\setlength{\abovecaptionskip}{0pt}
\centerline{
\hspace{-1.5cm}\subfigure{ \epsfxsize=15cm \epsffile{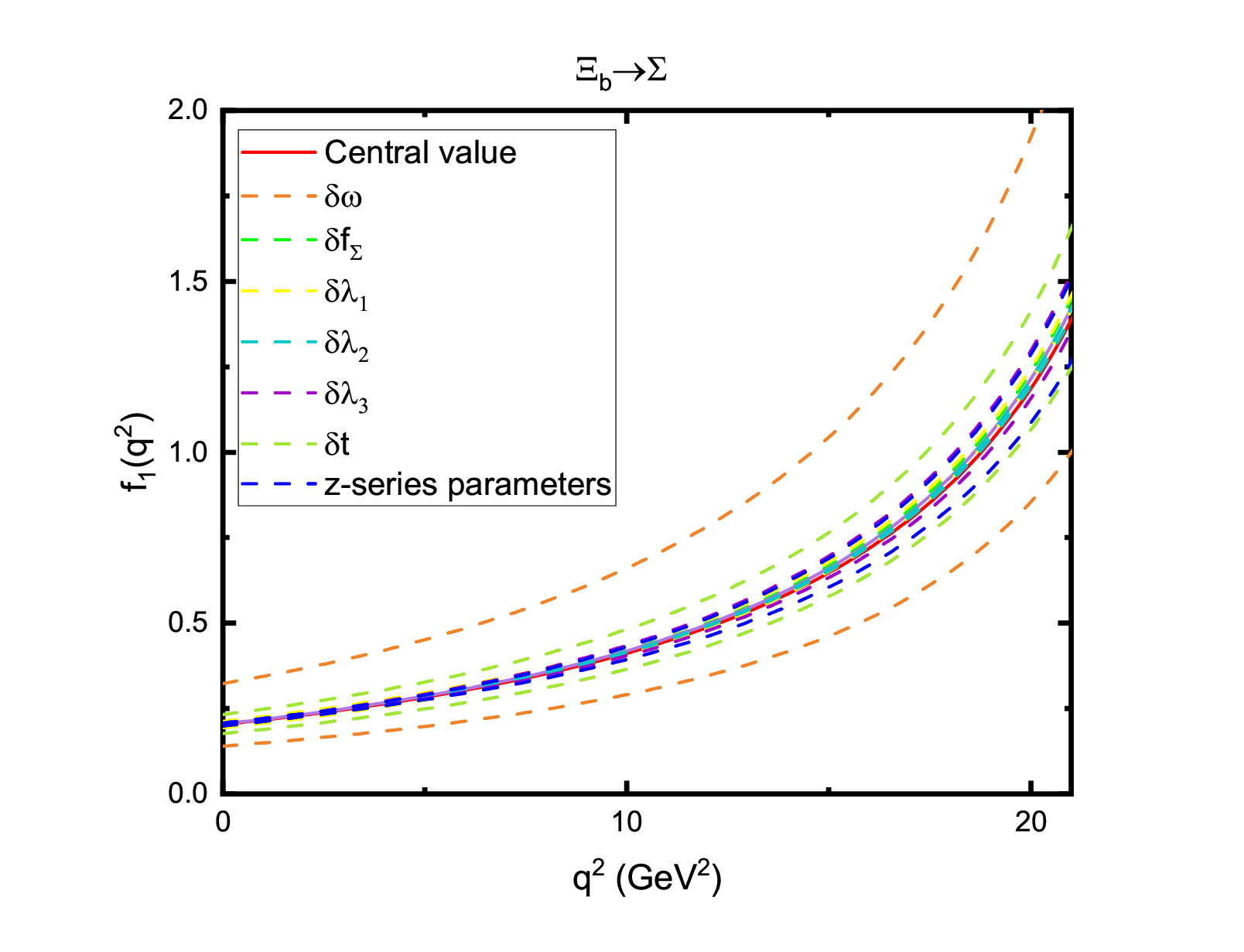}}}
\vspace{0.5cm}
\caption{
The theoretical uncertainties in the S1 scheme for the form factor $f_1(q^2)$ of $\Xi_b\rightarrow\Sigma$ transition.
The solid red line shows the $q^2$ shape at central values of model parameters.
The dashed orange curves show the effect of the variation of  $\omega_0=(0.4\pm0.1) \text{GeV}$ from the $\Xi_b$ baryon LCDAs.
The dashed green, yellow, cyan, and purple curves show the variation of the parameters $f_{\Sigma}$, $\lambda_1$, $\lambda_2$, and $\lambda_3$ of $\Sigma$ baryon LCDAs, respectively.
The dashed olive curves represent the scale $t$ variation between $0.8t$ and $1.2t$.
The uncertainty arising from the standard errors of the $z$-series parameters is displayed by the dashed blue curves.  }
 \label{fig:f1}
\end{center}
\end{figure}
The numerical values of the form factors at the maximum recoil point $q^2=0$ as well as the fitted parameters obtained from four different schemes are exhibited in Tables~\ref{tab:form1} and~\ref{tab:form2} for $\Xi_b^0\to \Sigma^+$ and $\Xi_b^-\to \Lambda$ transitions, respectively. There are three theoretical uncertainties in our calculations. The first quoted uncertainty is due to the nonperturbative parameters in the $\Xi_b$ baryonic LCDAs as shown in Eq.~(\ref{eq:mod}). The second uncertainty is caused by the variation of the parameters in the LCDAs of final baryons (see Eqs.~(\ref{eq:sigma}) and (\ref{eq:lambda})). The last one is from the hard scale $t$ varying from 0.8$t$ to 1.2$t$, which characterizes the size of higher-order corrections to the hard amplitudes.
 In Fig.~\ref{fig:f1}, we compare the relative importance of different sources of uncertainty in the form factors (take the $f_1(q^2)$ of $\Xi_b\rightarrow\Sigma$ transition as an example). The solid red curve denotes the shape that all the input parameters are set as their central values. Varying each parameter in a given range (see Eqs.(\ref{eq:mod})-(\ref{eq:lambda})) and redoing the above fitting procedure yields two curves with the same color as shown in Fig.~\ref{fig:f1},  which are regarded as the upper and lower bounds of the uncertainty.
In this case,  PQCD results at different $q^2$ values are correlations.
The uncertainty  due to the standard errors of the $z$-series parameters (shown by the dashed blue curves), derived from the covariance matrix of the NonlinearModelFit, is typically smaller than the uncertainties of the input parameters and is therefore neglected in our analysis. The slight effect from the $z$-expansion uncertainty demonstrates the excellent fit quality of the chosen model. The numerical values of different sources of uncertainties for the form factors at the large recoil are presented in Tables.~\ref{tab:form1} and~\ref{tab:form2}, where the uncertainties from several independent parameters in the LCDAs of final baryons have been combined in quadrature.  
It turns out that the values in S1 and S4 exhibit greater uncertainties than those in S2 and S3. The primary reason can be attributed to the strong parameter dependence in the Exponential and Free parton models for the $\Xi_b$ baryon LCDAs. Any significant reduction of the error requires more accurate information on the LCDAs. 
 Furthermore, the authors in~\cite{Blake:2022vfl} have proposed a novel parametrization of $\Lambda_b\to \Lambda$ form factors using orthonormal polynomials that diagonalizes the form factor contributions to the dispersive bounds, offering excellent control of systematic uncertainties when extrapolating from low to large hadronic recoil. 
This improvement imposes additional unitarity constraints on
the  $z$-series parameters,  
hopefully reducing the theoretical uncertainty of  form factors. It is worth exploring when the improved parametrization can be used to an extrapolation from large to low recoil in the future. 

We now compare these form factors with other  predictions existing in the literature. From Table~\ref{tab:form1}, it is evident that the results from S1, S2, and S3 schemes are basically consistent with each other and close to those obtained in~\cite{Zhao:2018zcb}, while those from the S4 scheme are slightly large in magnitude. From Table~\ref{tab:form2}, we learn that the solutions from  S1 and S2 are nearly identical, while those from  S3 and S4 are of similar sizes. Although the magnitudes of the form factors from different approaches are in agreement within the error bars, there exist some uncertainties about the sign of these form factors among different literature. All our four solutions show $f_1,g_1,f_2, g_2>0$ and $f_3,g_3<0$, which are the same as those derived by the relativistic quark-diquark model in Refs.~\cite{Faustov:2018ahb, Zhang:2019xdm}. The minus signs of $f_1(0)$ and $g_1(0)$ from~\cite{Zhao:2018zcb} in Table~\ref{tab:form2} are attributed to the negative values of the corresponding overlap factor. The obtained values of $f_2(0)$ and $g_2(0)$ from~\cite{Zhao:2018zcb} in Table~\ref{tab:form1} are also opposite in sign with ours. More careful analysis should be performed to fix this problem. From these tables, we make a general observation: these form factors satisfy the heavy quark effective theory relations, $f_1(0)\approx g_1(0)$ and $ f_2(0)\approx g_2(0)\approx f_3(0)\approx g_3(0)\approx 0$.  

\begin{figure}[!htbh]
\begin{center}
\setlength{\abovecaptionskip}{0pt}
\centerline{
\hspace{-1.0cm}\subfigure{ \epsfxsize=5cm \epsffile{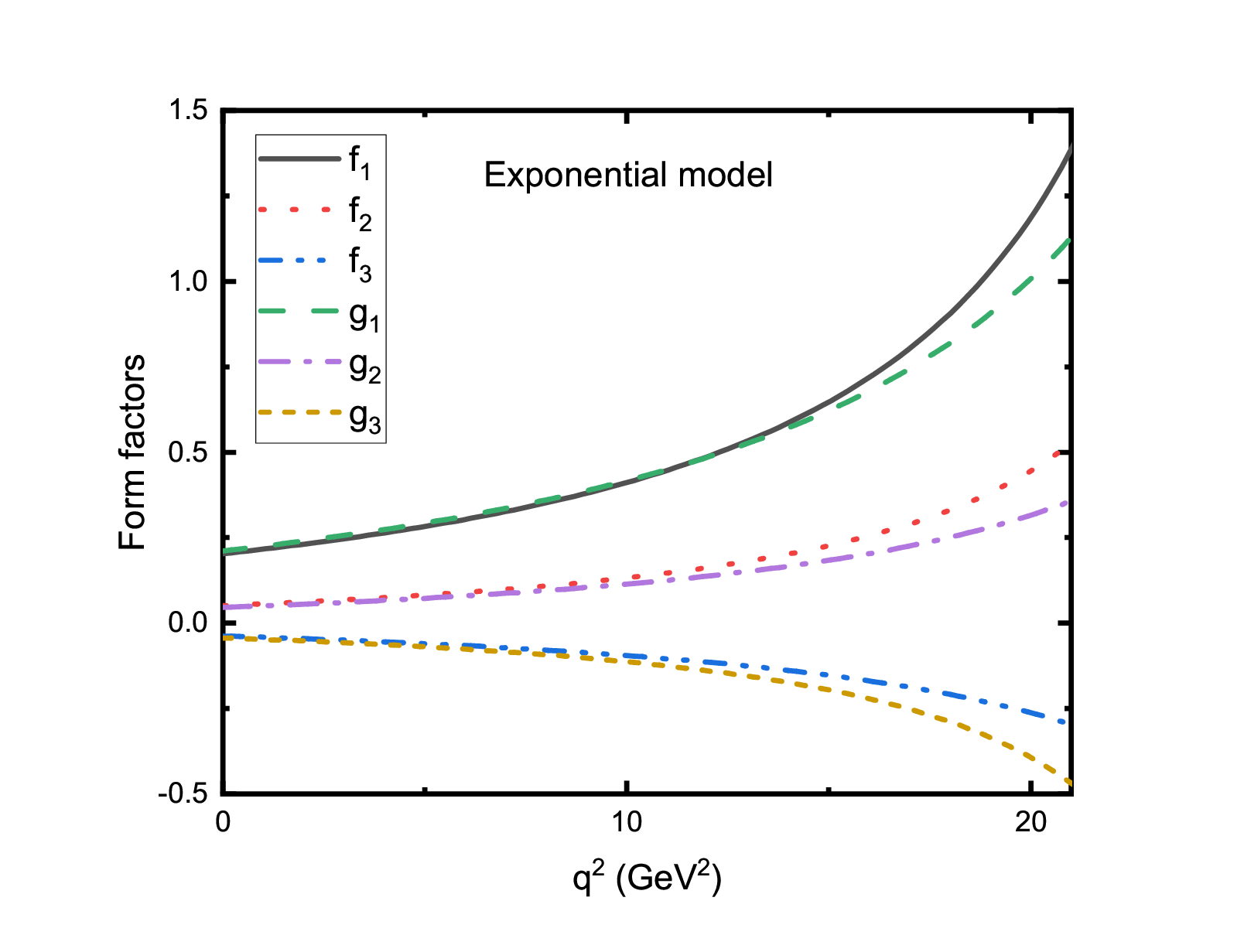}}
\hspace{-1.0cm}\subfigure{ \epsfxsize=5cm \epsffile{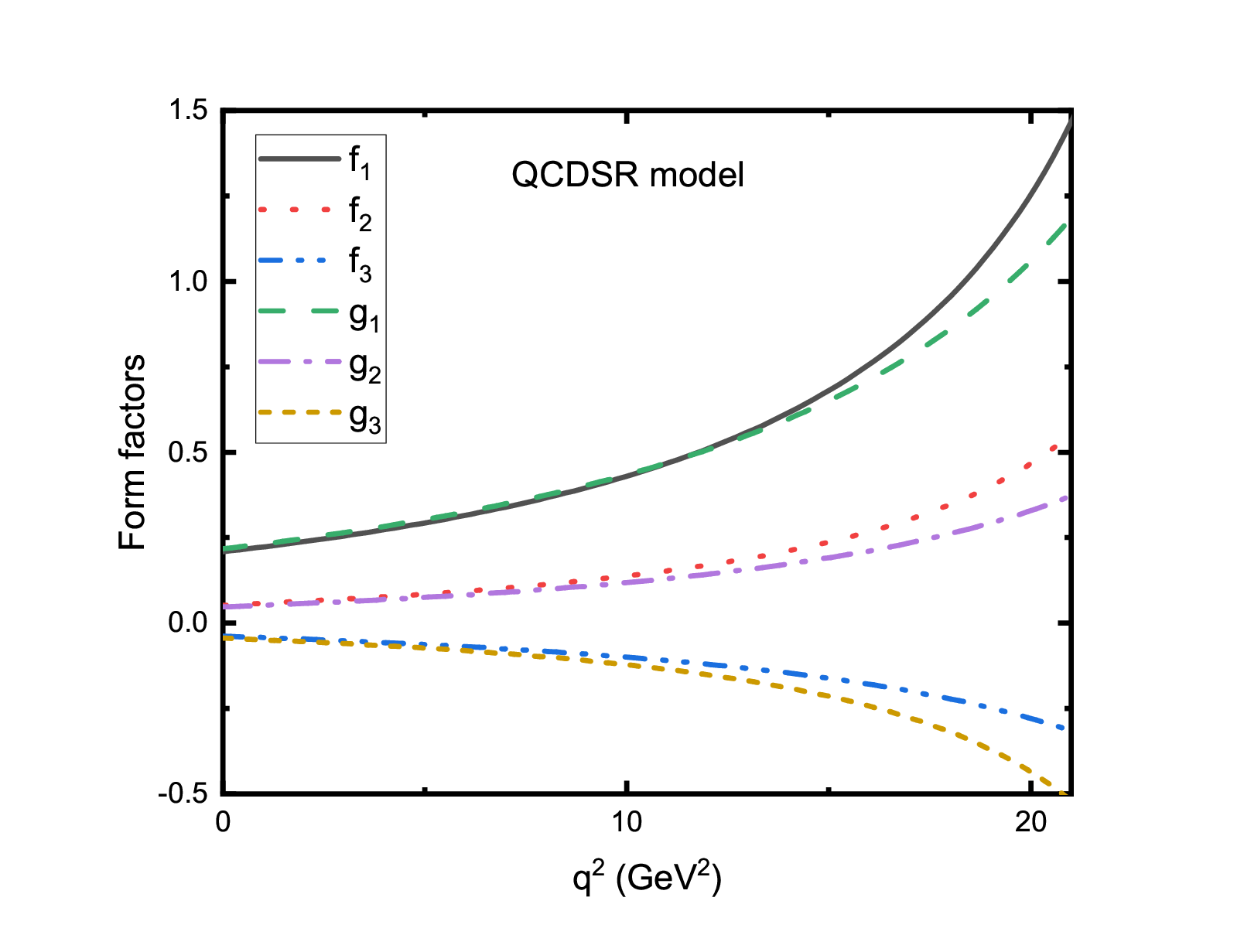}}
\hspace{-1.0cm}\subfigure{ \epsfxsize=5cm \epsffile{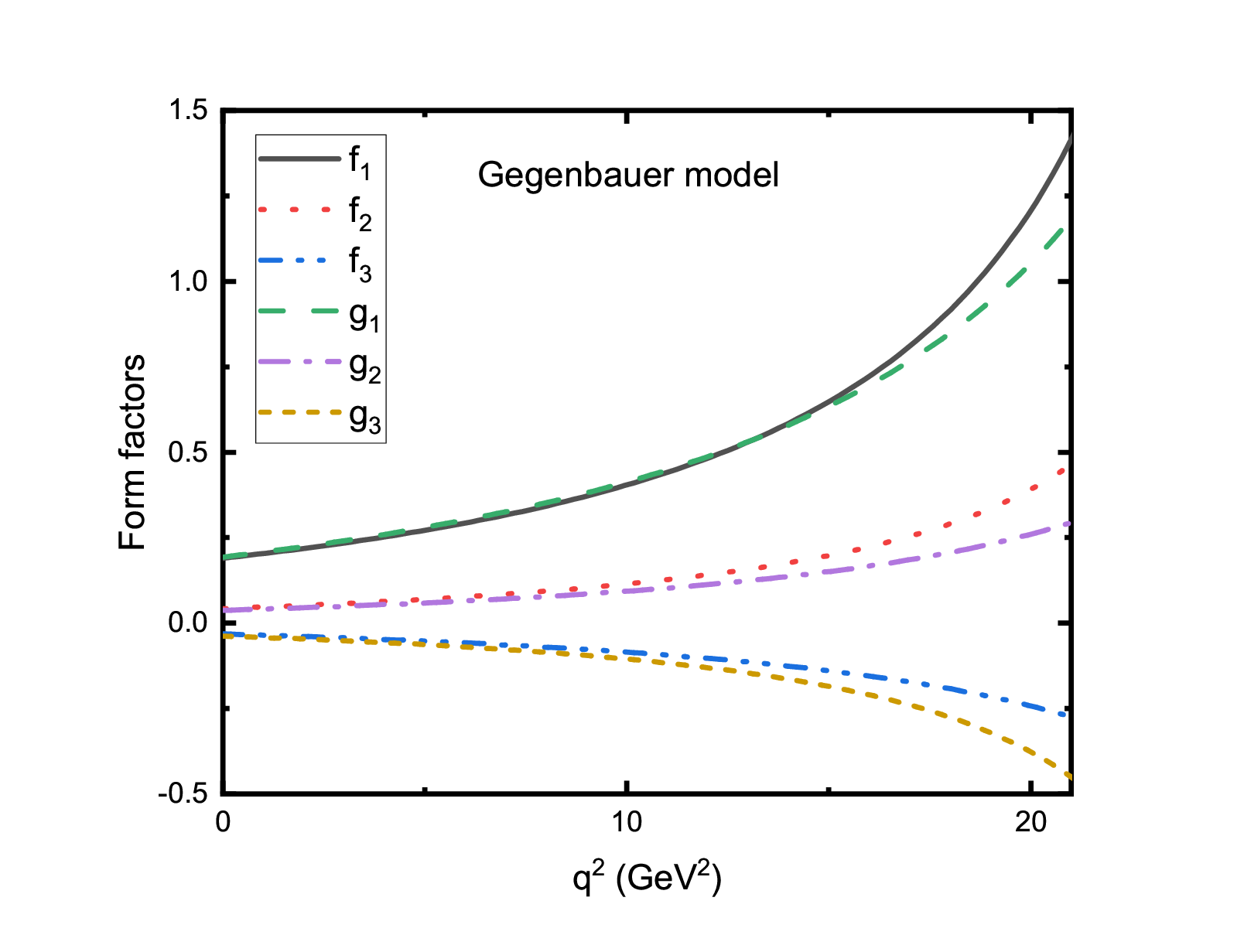}}
\hspace{-1.0cm}\subfigure{ \epsfxsize=5cm \epsffile{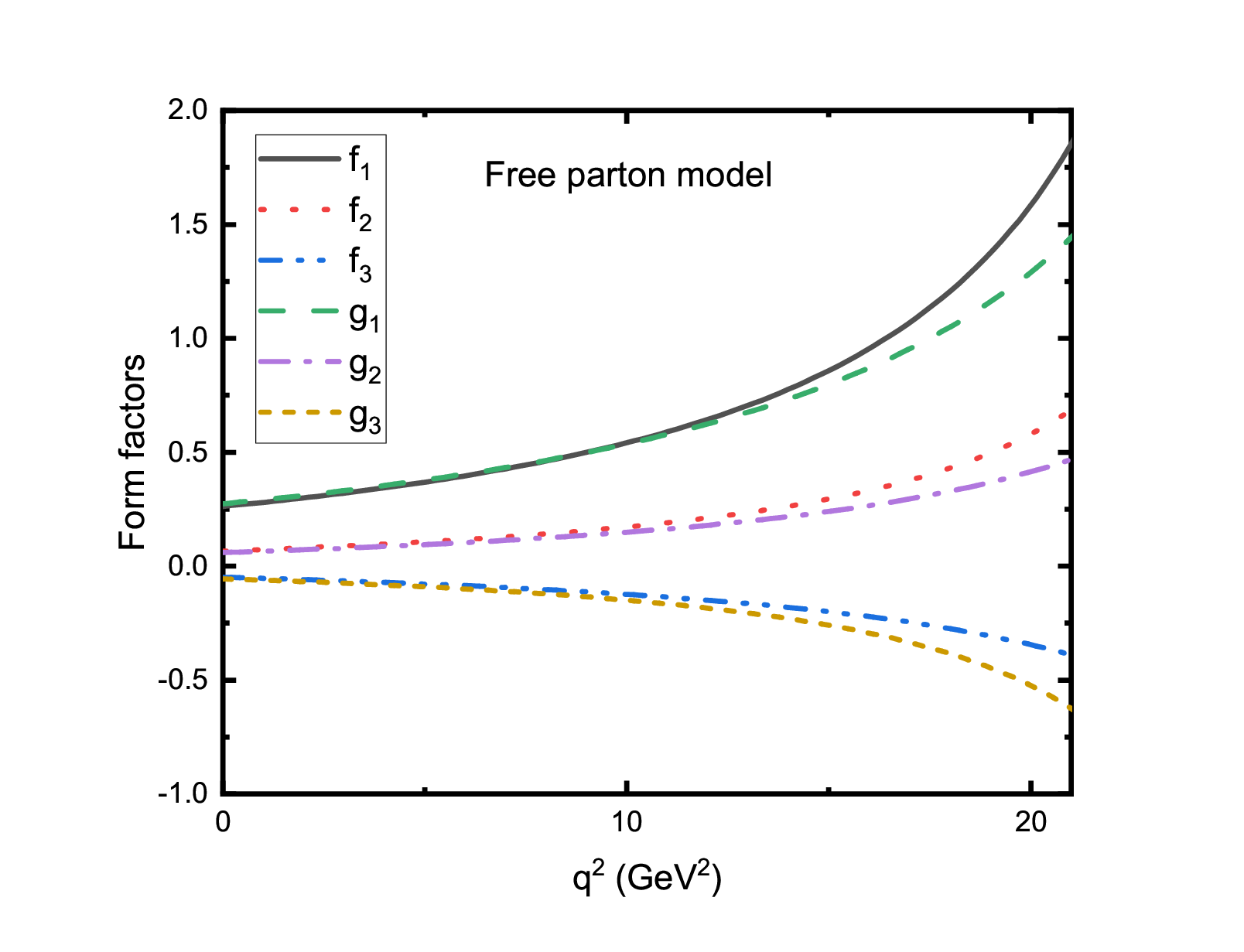}}}
\vspace{1cm}
\centerline{
\hspace{-1.0cm}\subfigure{ \epsfxsize=5cm \epsffile{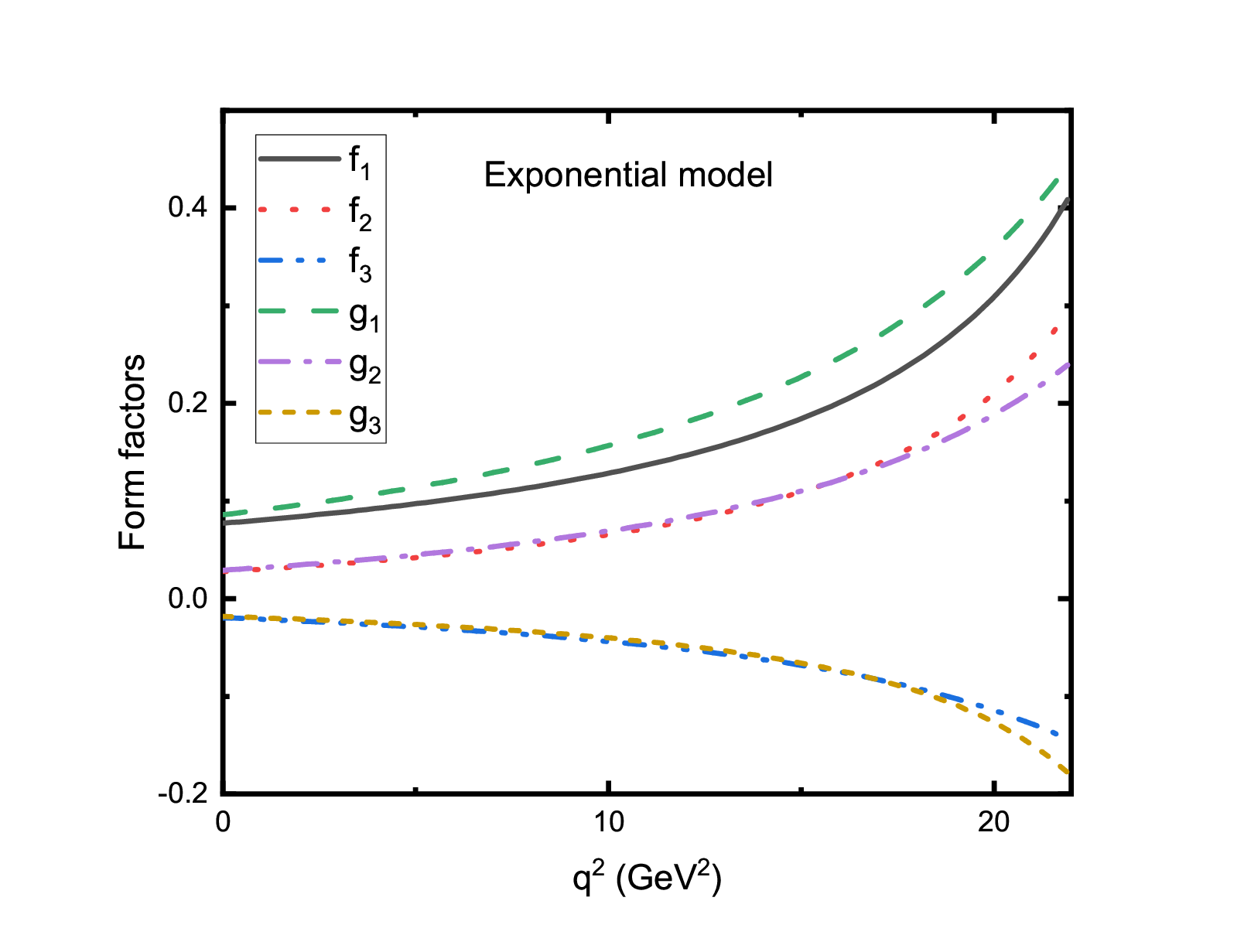}}
\hspace{-1.0cm}\subfigure{ \epsfxsize=5cm \epsffile{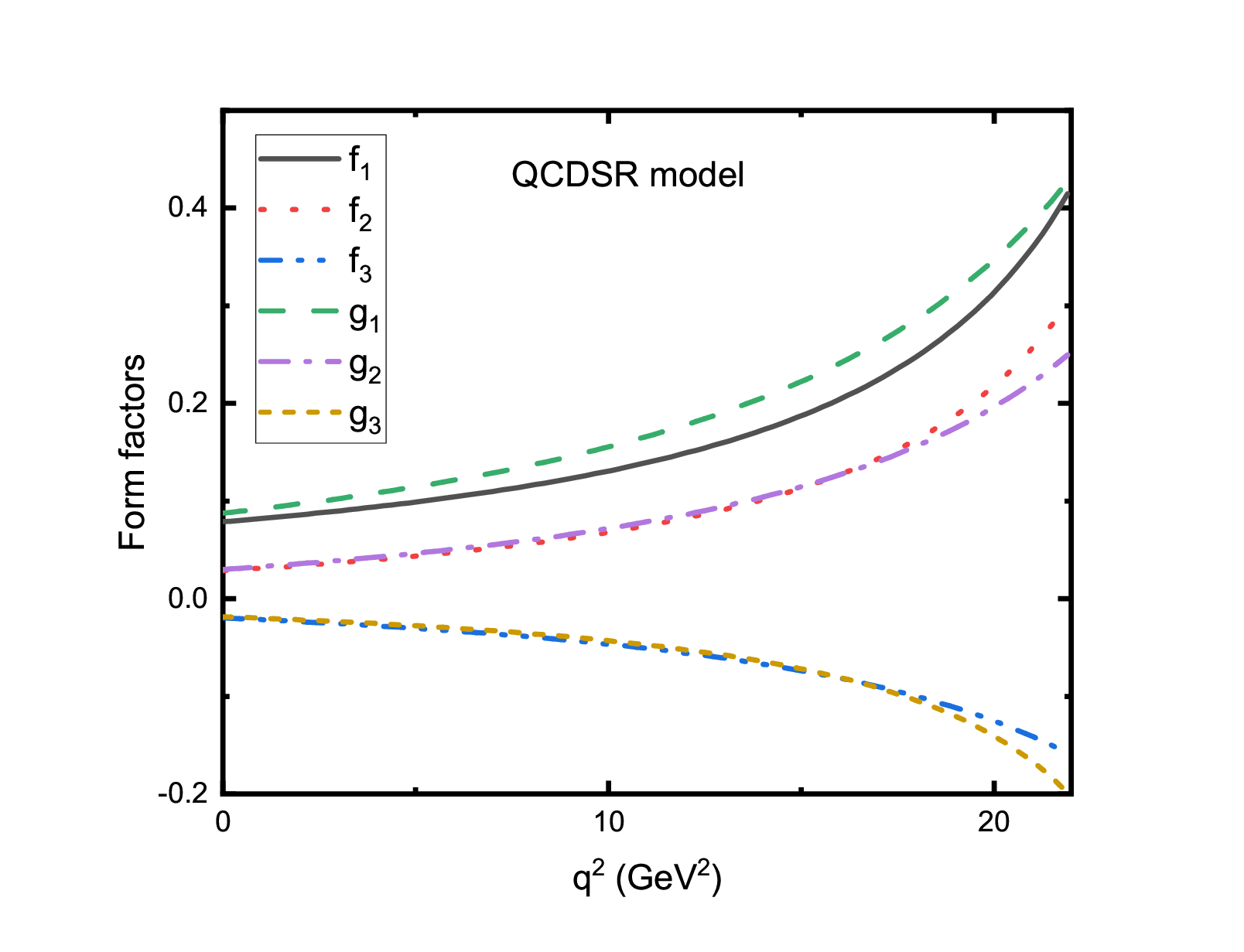}}
\hspace{-1.0cm}\subfigure{ \epsfxsize=5cm \epsffile{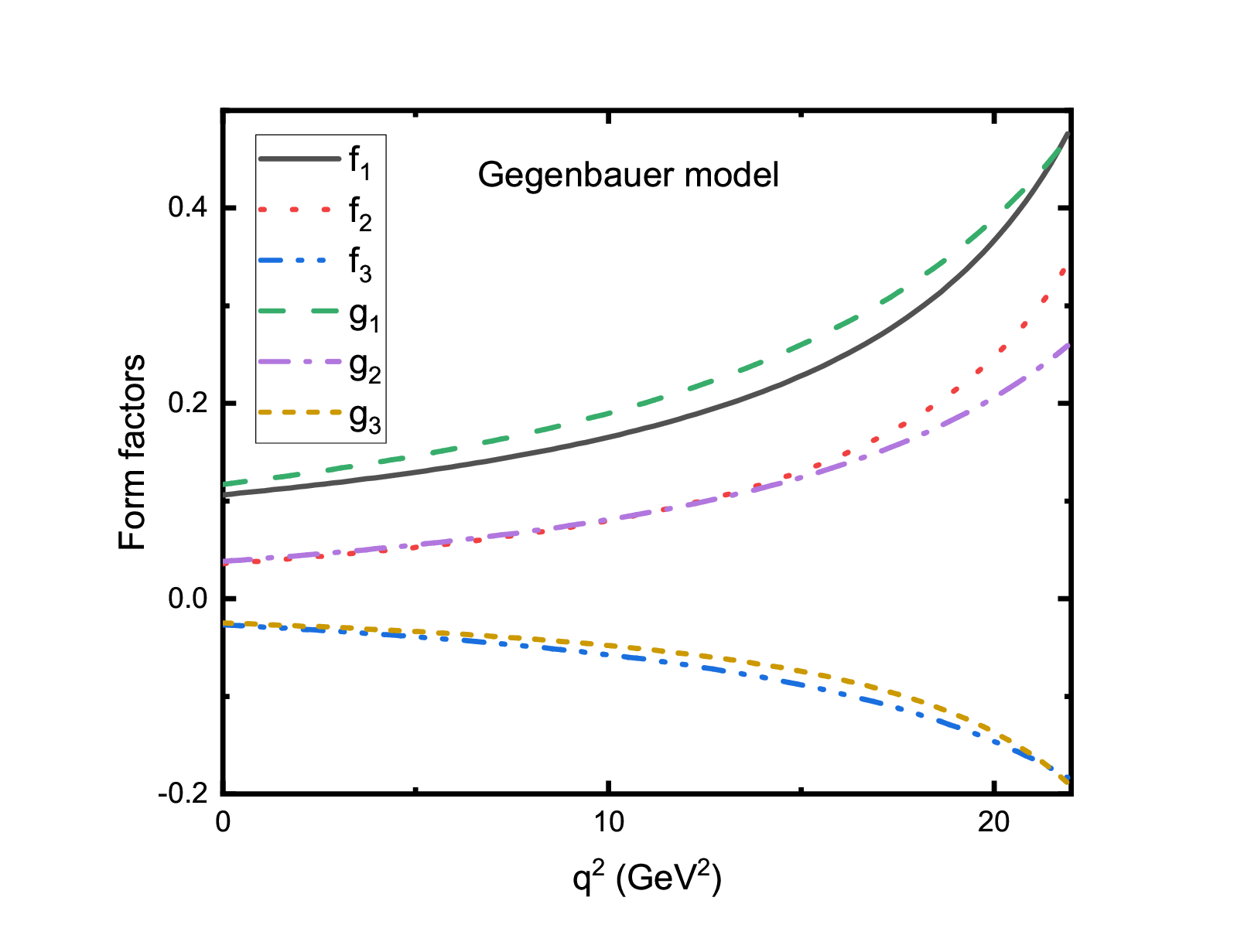}}
\hspace{-1.0cm}\subfigure{ \epsfxsize=5cm \epsffile{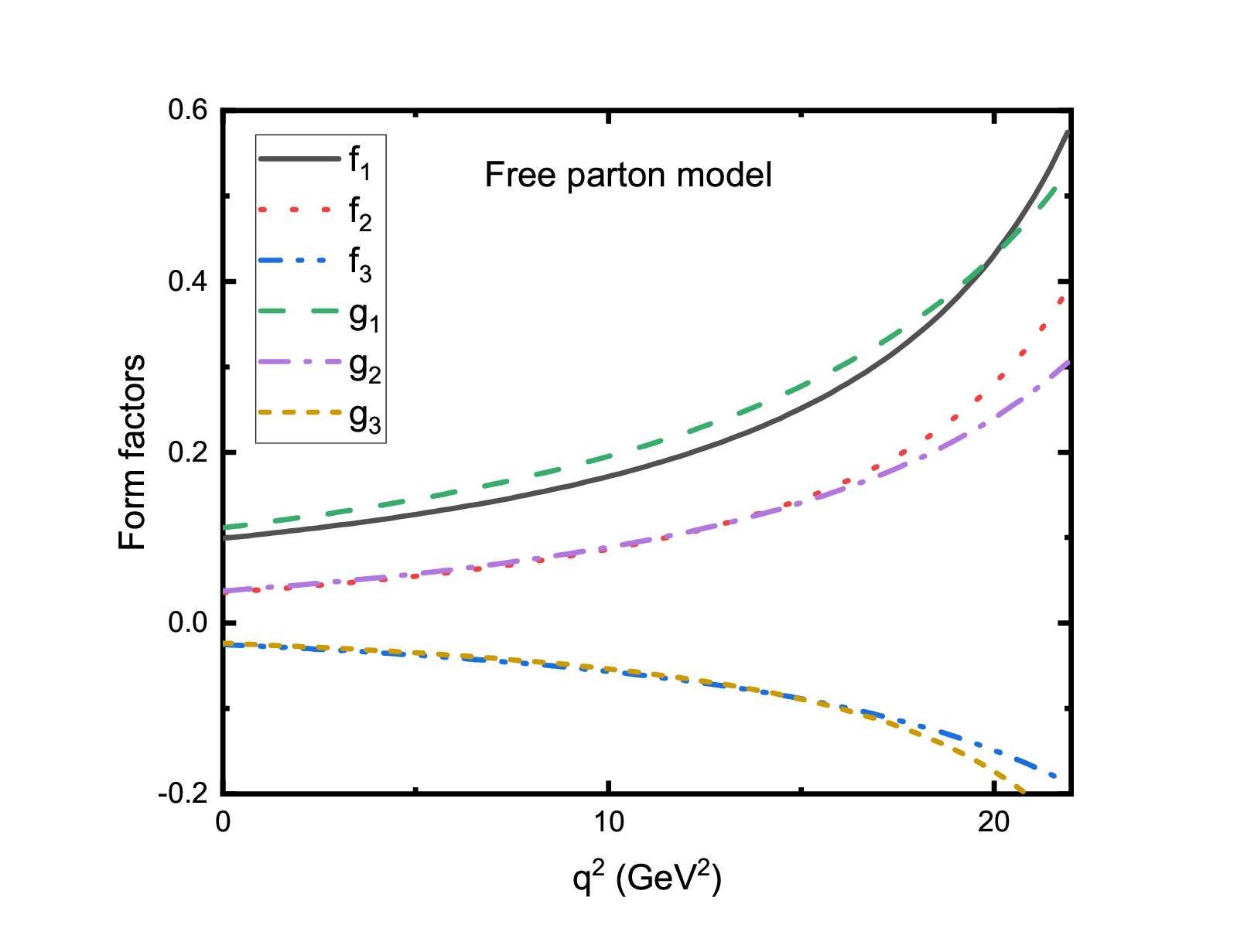}}}
\vspace{1cm}
\caption{$q^2$ dependencies of the form factors with four different models of $\Xi_b$ baryonic LCDAs. The solid (black),  dashed (green), dotted (red), dot-dashed (magenta), double dot-dashed (blue), and the short dashed (orange) lines denote the $f_1$, $g_1$, $f_2$, $g_2$, $f_3$, and $g_3$, respectively. Upper and lower tows correspond to $\Xi_b^0\to \Sigma^+$ and $\Xi_b^-\to \Lambda$ transitions, respectively.}
 \label{fig:Form}
\end{center}
\end{figure}

Because the form factors are extrapolated across the entire physical region, we plot the dependence of these form factors on $q^2$ in their allowed regions in Fig.~\ref{fig:Form}. The shapes obtained by the Exponential, QCDSR, Gegenbauer, and Free parton models are displayed sequentially from left to right. Upper and lower panels correspond to $\Xi_b^0\to \Sigma^+$ and $\Xi_b^-\to \Lambda$ transitions, respectively. As it is expected from weak decays,   the magnitude of each form factor steadily increases with increasing $q^2$. The dominant form factors are $f_1$ and $g_1$, whereas the remaining four are much smaller in size and show good stability with respect to the variations of $q^2$. $f_i(q^2)$ and $g_i(q^2)$ possess a shared $q^2$ shape in the low $q^2$ area, but separate each other in the high $q^2$ region. Once more, we observe that $f_1,g_1,f_2, g_2$ take positive values while $f_3,g_3$ take negative ones over the entire $q^2$ range for all the four solutions. The above observed patterns are similar to those of $\Lambda_b\to p$ transition obtained in~\cite{Gutsche:2014zna}.

\begin{table}[!htbh]
	\caption{Comparison of theoretical predictions on the branching ratios ($10^{-4}$),  the LFU ratio, and integrated angular observables for the semileptonic $\Xi_b^0\to \Sigma^+ \ell \nu_\ell$ decay. }
	\label{tab:branching1}
	\begin{tabular}[t]{lccccc}
	\hline\hline
          Observable           & This work (S1)   & This work (S2)   &This work (S3)    &This work (S4)                &~\cite{Zhao:2018zcb}\\ \hline
    $\mathcal{B}_\tau$         &$4.3^{+4.3}_{-1.5}$     &$4.8^{+1.3}_{-0.8}$     &$4.6^{+0.9}_{-0.4}$ &$7.3^{+5.6}_{-2.6}$ &$\cdots$ \\
    $\mathcal{B}_e$            &$6.8^{+6.8}_{-2.4}$     &$7.4^{+1.9}_{-1.3}$     &$7.0^{+1.6}_{-0.8}$ &$11.3^{+11.9}_{-4.0}$ &2.83\\
    $\mathcal{R}_{\Sigma}$      &$0.645^{+0.011}_{-0.002}$    &$0.648^{+0.005}_{-0.006}$    &$0.660^{+0.014}_{-0.008}$&$0.647^{+0.008}_{-0.002}$   &$\cdots$  \\
    $\langle A^\tau_{\text{FB}}  \rangle$  &$0.053^{+0.008}_{-0.007}$&$0.055^{+0.003}_{-0.005}$&$0.063^{+0.005}_{-0.003}$&$0.052^{+0.006}_{-0.010}$&$\cdots$ \\
    $\langle A^e_{\text{FB}}     \rangle$  &$0.315^{+0.003}_{-0.002}$&$0.316^{+0.000}_{-0.004}$&$0.313^{+0.008}_{-0.004}$&$0.316^{+0.002}_{-0.005}$&$\cdots$ \\
    $\langle C^\tau_{\text{F}}\rangle $      &$-0.193^{+0.006}_{-0.003}$&$-0.193^{+0.003}_{-0.004}$&$-0.187^{+0.004}_{-0.006}$&$-0.195^{+0.002}_{-0.004}$&$\cdots$ \\
    $\langle C^e_{\text{F}}\rangle $         &$-0.499^{+0.018}_{-0.015}$&$-0.495^{+0.014}_{-0.015}$&$-0.466^{+0.011}_{-0.013}$&$-0.504^{+0.007}_{-0.018}$&$\cdots$ \\
    $\langle H_{z}^{\tau}\rangle$            &$-0.967^{+0.009}_{-0.003}$&$-0.967^{+0.011}_{-0.004}$&$-0.947^{+0.008}_{-0.015}$&$-0.971^{+0.002}_{-0.001}$&$\cdots$ \\
    $\langle H_{z}^{e}\rangle$               &$-0.967^{+0.010}_{-0.009}$&$-0.967^{+0.013}_{-0.005}$&$-0.943^{+0.009}_{-0.018}$&$-0.973^{+0.006}_{-0.002}$&$\cdots$ \\
    $\langle H_{x}^{\tau}\rangle$            &$-0.071^{+0.018}_{-0.020}$&$-0.071^{+0.020}_{-0.029}$&$-0.115^{+0.031}_{-0.025}$&$-0.054^{+0.016}_{-0.012}$&$\cdots$ \\
    $\langle H_{x}^{e}\rangle$               &$-0.127^{+0.006}_{-0.021}$&$-0.128^{+0.028}_{-0.025}$&$-0.195^{+0.057}_{-0.028}$&$-0.105^{+0.010}_{-0.016}$&$\cdots$ \\
    $\langle P_{z}^{\tau}\rangle$            &$-0.414^{+0.018}_{-0.015}$&$-0.417^{+0.013}_{-0.005}$&$-0.438^{+0.018}_{-0.011}$&$-0.409^{+0.015}_{-0.005}$&$\cdots$ \\
    $\langle P_{x}^{\tau}\rangle$            &$0.672^{+0.021}_{-0.010}$&$0.671^{+0.008}_{-0.008}$  &$0.652^{+0.015}_{-0.007}$&$0.676^{+0.007}_{-0.005}$&$\cdots$ \\
		\hline\hline
	\end{tabular}
\end{table}

\begin{table}[!htbh]
	\caption{Same as Table~\ref{tab:branching1} but for  $\Xi_b^-\to \Lambda \ell \nu_\ell$. The branching ratios are given in the unit of $10^{-5}$. }
	\label{tab:branching2}
	\begin{tabular}[t]{lcccccc}
	\hline\hline
          Observable           & This work (S1)   & This work (S2)   &This work (S3)    &This work (S4)                &~\cite{Zhao:2018zcb}
          &~\cite{Faustov:2018ahb}\\ \hline
    $\mathcal{B}_\tau$         &$5.3^{+6.1}_{-2.8}$     &$5.1^{+2.6}_{-1.9}$     &$7.5^{+1.8}_{-2.3}$ &$8.5^{+10.5}_{-4.1}$   &$\cdots$ &$18$\\
    $\mathcal{B}_e$            &$8.5^{+9.6}_{-4.3}$     &$8.4^{+4.0}_{-3.0}$     &$12.8^{+2.8}_{-3.2}$   &$13.8^{+16.7}_{-7.1}$  &$5.42$  &$26$\\
    $\mathcal{R}_{\Lambda}$      &$0.617^{+0.002}_{-0.004}$    &$0.611^{+0.011}_{-0.009}$    &$0.588^{+0.003}_{-0.008}$ &$0.616^{+0.011}_{-0.010}$&$\cdots$ &$0.717\pm 0.021$\\
    $\langle A^\tau_{\text{FB}}  \rangle$  &$0.044^{+0.003}_{-0.005}$&$0.036^{+0.000}_{-0.007}$&$0.025^{+0.004}_{-0.003}$&$0.037^{+0.010}_{-0.004}$&$\cdots$ &0.213 \\
    $\langle A^e_{\text{FB}}     \rangle$  &$0.313^{+0.000}_{-0.006}$&$0.309^{+0.001}_{-0.003}$&$0.302^{+0.001}_{-0.006}$&$0.310^{+0.007}_{-0.004}$&$\cdots$ &0.384\\
    $\langle C^\tau_{\text{F}}\rangle $      &$-0.198^{+0.003}_{-0.002}$&$-0.200^{+0.001}_{-0.004}$&$-0.200^{+0.005}_{-0.003}$&$-0.199^{+0.007}_{-0.002}$&$\cdots$ &$-0.073$\\
    $\langle C^e_{\text{F}}\rangle $         &$-0.540^{+0.010}_{-0.009}$&$-0.553^{+0.005}_{-0.011}$&$-0.578^{+0.007}_{-0.009}$&$-0.549^{+0.025}_{-0.002}$&$\cdots$ &$-0.226$\\
    $\langle H_{z}^{\tau}\rangle$            &$-0.983^{+0.012}_{-0.000}$&$-0.987^{+0.005}_{-0.002}$&$-0.989^{+0.007}_{-0.005}$&$-0.988^{+0.015}_{-0.001}$&$\cdots$ &$-0.903$\\
    $\langle H_{z}^{e}\rangle$               &$-0.984^{+0.014}_{-0.001}$&$-0.988^{+0.004}_{-0.007}$&$-0.990^{+0.014}_{-0.004}$&$-0.989^{+0.016}_{-0.001}$&$\cdots$ &$-0.919$\\
    $\langle H_{x}^{\tau}\rangle$            &$-0.049^{+0.008}_{-0.031}$&$-0.015^{+0.037}_{-0.007}$&$-0.015^{+0.014}_{-0.052}$&$-0.012^{+0.015}_{-0.060}$&$\cdots$ &$\cdots$\\
    $\langle H_{x}^{e}\rangle$               &$-0.040^{+0.010}_{-0.052}$&$-0.010^{+0.044}_{-0.012}$&$-0.004^{+0.020}_{-0.052}$&$-0.040^{+0.021}_{-0.082}$&$\cdots$ &$\cdots$\\
    $\langle P_{z}^{\tau}\rangle$            &$-0.401^{+0.010}_{-0.010}$&$-0.390^{+0.011}_{-0.004}$&$-0.377^{+0.009}_{-0.011}$&$-0.389^{+0.000}_{-0.020}$&$\cdots$ &$-0.579$\\
    $\langle P_{x}^{\tau}\rangle$            &$0.686^{+0.002}_{-0.005}$&$0.691^{+0.008}_{-0.001}$&$0.695^{+0.007}_{-0.010}$&$0.692^{+0.001}_{-0.018}$&$\cdots$ &$\cdots$\\
		\hline\hline
	\end{tabular}
\end{table}
Through the above preparation, we now turn our attention to the calculations of the semileptonic decays observables for all lepton species. Since the electron and muon are very light compared with the heavy $\tau$ lepton, we neglect their masses in the numerical calculations. The predicted branching ratios and various integrated angular observables  for the $\Sigma$ and $\Lambda$ modes are summarized in Tables~\ref{tab:branching1}  and~\ref{tab:branching2}, respectively. The considered theoretical uncertainties have the same meaning as in Table~\ref{tab:form1} but are combined in quadrature. The branching ratios from S1 and S4 show greater sensitivity to variations due to uncertainties corresponding to the model parameters in the $\Xi_b$ LCDAs. Their overall uncertainties can overcome $100\% $ in magnitude. The total uncertainties of the branching ratios from S3 are less than $30\% $. The predictions on the branching ratios can further be improved by more precise determination of the input parameters appearing in LCDAs of  baryons  as well as taking into account higher-power corrections. The obtained branching ratios for the $\Sigma(\Lambda)$ mode are of the order of $10^{-4(-5)}$, indicating promise for measurement in future experiments.

As expected, the decays involving heavier leptons are kinematically suppressed. To further test the lepton universality, we compute the ratio  $\mathcal{R}_{\mathcal{B}_f}=\frac{\mathcal{B}_\tau}{\mathcal{B}_e}$ (see Tables~\ref{tab:branching1} and~\ref{tab:branching2}), which is one of the most important probes to search for new physics effects. It is clear that the predicted $\mathcal{R}_{\mathcal{B}_f}$ from different models are generally around 0.6. These values are compatible with the SM expectations of $\mathcal{R}_{\pi}=0.641\pm0.016$~\cite{Bernlochner:2021vlv,Bernlochner:2015mya} and $\mathcal{R}_{p}=0.688\pm0.064$~\cite{Duan:2024ayo}, but larger than the counterparts in the $b \to c $  sector, such as $\mathcal{R}_{D^{(*)}}$~\cite{Bernlochner:2021vlv,Cui:2023jiw}  and $\mathcal{R}_{\Lambda_c(\Xi_c)}$ \cite{Bernlochner:2018bfn, Detmold:2015aaa, Bernlochner:2018kxh, Fedele:2022iib,Rui:2025iwa}. This observed pattern  reflects the substantial contribution of the    spin-flip component in the $b\to u$ transition. With the successful running of Belle II and LHCb Run-3, more and more semileptonic  $b\to u$ processes will be measured. Measurements of these ratios offer an orthogonal path to probe the anomalies observed in $\mathcal{R}_{D^{(*)}}$. Future experimental data will tell us whether there are LFU violations in these channels or not.

From Tables~\ref{tab:branching1} and~\ref{tab:branching2}, it is evident that the PQCD predictions on the integrated asymmetries are not sensitive to the different models of $\Xi_b$ LCDAs. These quantities are also  insensitive to the above considered theoretical uncertainties because of the partial cancellation in the ratios of the helicity structure. Apart from the longitudinal hadronic polarizations, all other asymmetry observables of decays involving $\tau$  show significant differences compared to electron modes, primarily due to the obvious phase space difference between tauonic and electron channels. These results can be checked experimentally.

In Tables~\ref{tab:branching1} and~\ref{tab:branching2}, we also provide a comparison with results obtained within other approaches~\cite{Zhao:2018zcb,Faustov:2018ahb}. In~\cite{Zhao:2018zcb}, the authors analyzed in detail the semileptonic  weak decays of the heavy baryons in the light-front approach. Only the branching ratios for the electronic modes are provided in~\cite{Zhao:2018zcb}. Our branching ratios from S1 and S2 for the $\Lambda$ channel are consistent with their values within uncertainty bands, while the central values for the $\Sigma$ mode are larger by a few factors. In~\cite{Faustov:2018ahb}, the $\Xi_b^-\to \Lambda \ell \nu_\ell$ decays were calculated in the relativistic quark-diquark picture with the comprehensive account of the relativistic effects. Apart from the branching ratios, various asymmetries were also predicted as shown in Table~\ref{tab:branching2}. Although their branching ratios are several times bigger, the relative ratio  $\mathcal{R}_{\Lambda}$  is  consistent with ours. In general, the obtained asymmetries in~\cite{Faustov:2018ahb} are comparable with ours except that its $\langle A_{\text{FB}}\rangle$($\langle C_{\text{F}}\rangle $) is large (small) in size.

\begin{figure}[!htbh]
\begin{center}
\setlength{\abovecaptionskip}{0pt}
\centerline{
\hspace{-1.0cm}\subfigure{ \epsfxsize=8cm \epsffile{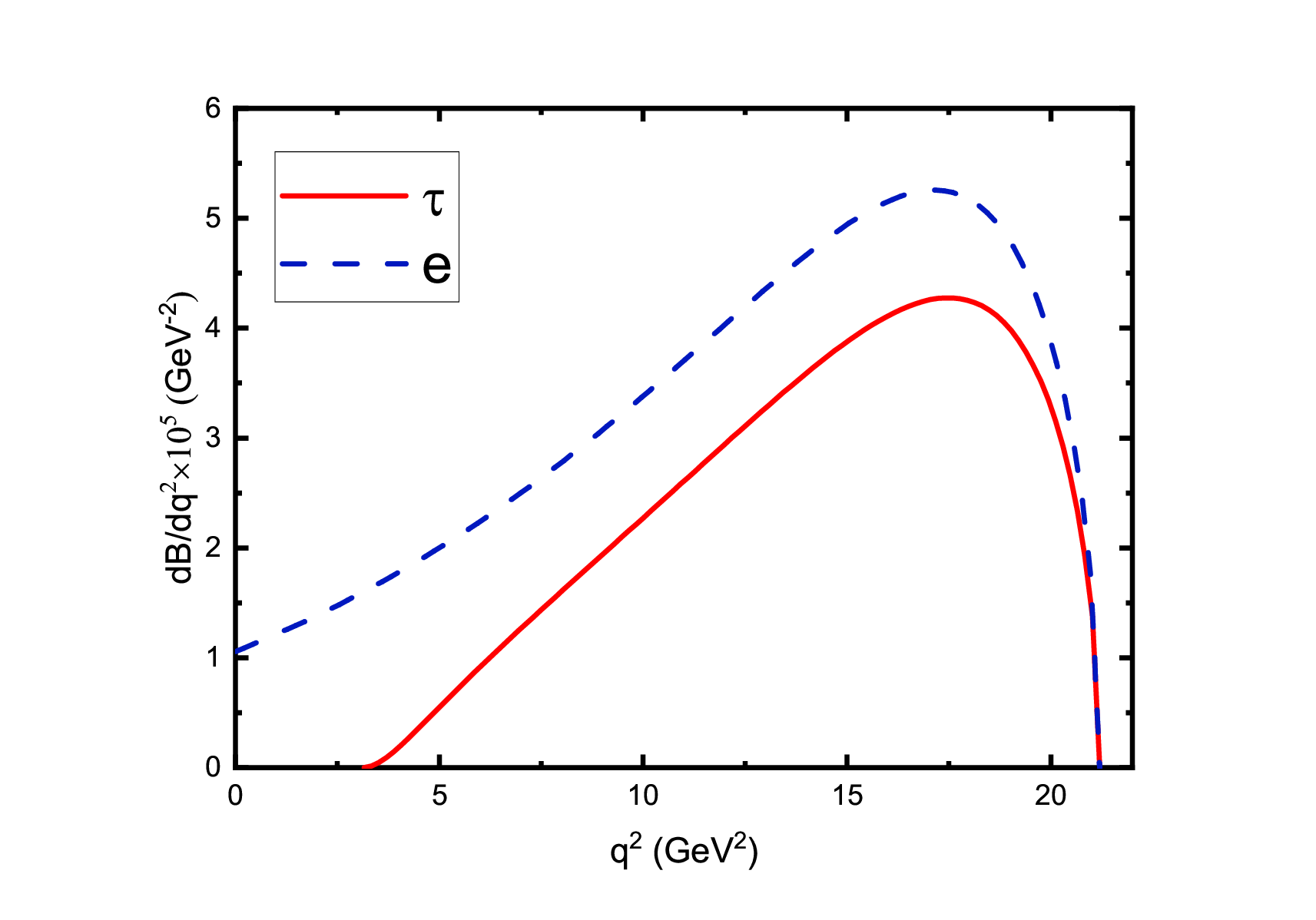}}
\hspace{-1.0cm}\subfigure{ \epsfxsize=8cm \epsffile{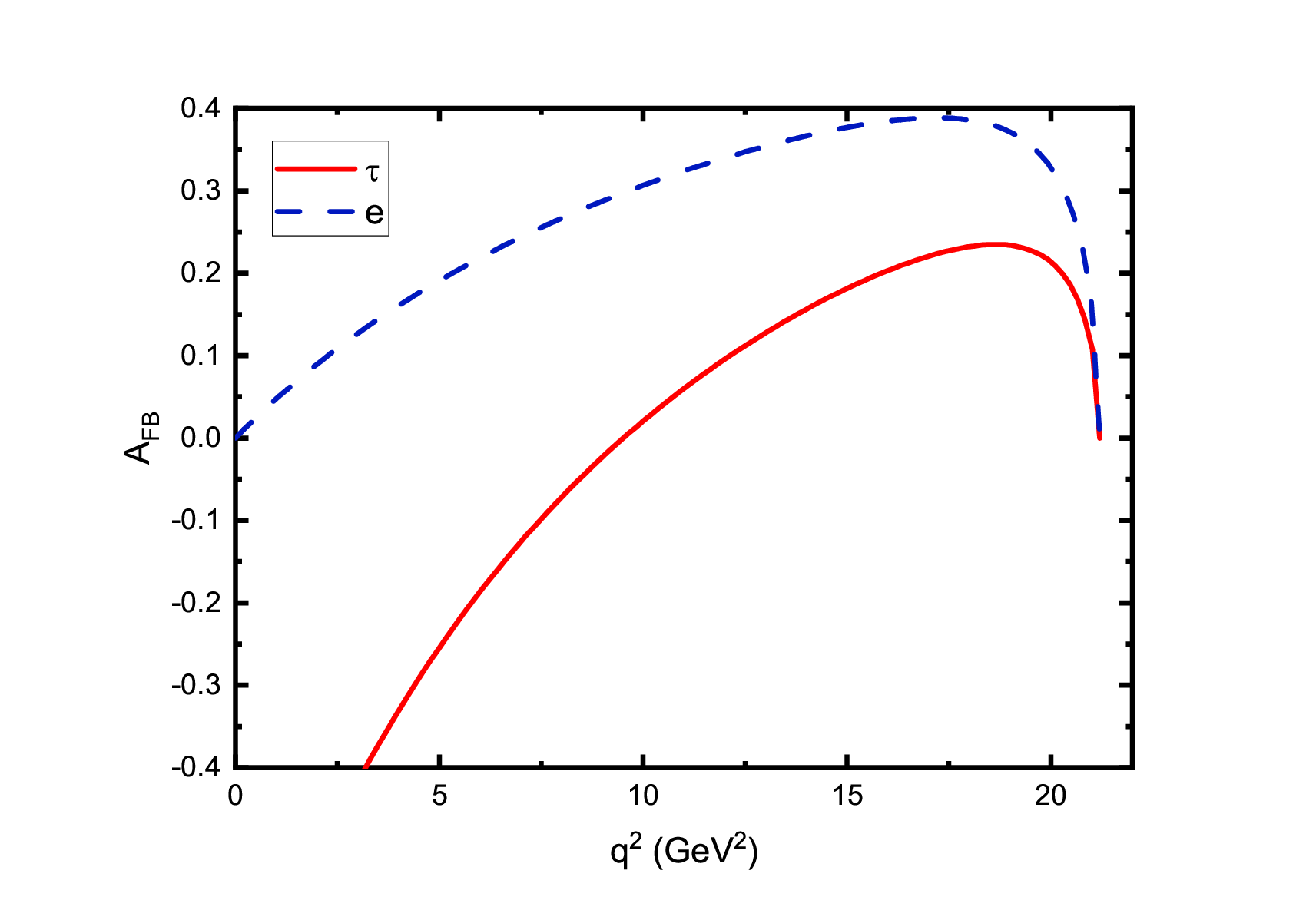}}}
\vspace{1cm}
\centerline{
\hspace{-1.0cm}\subfigure{ \epsfxsize=8cm \epsffile{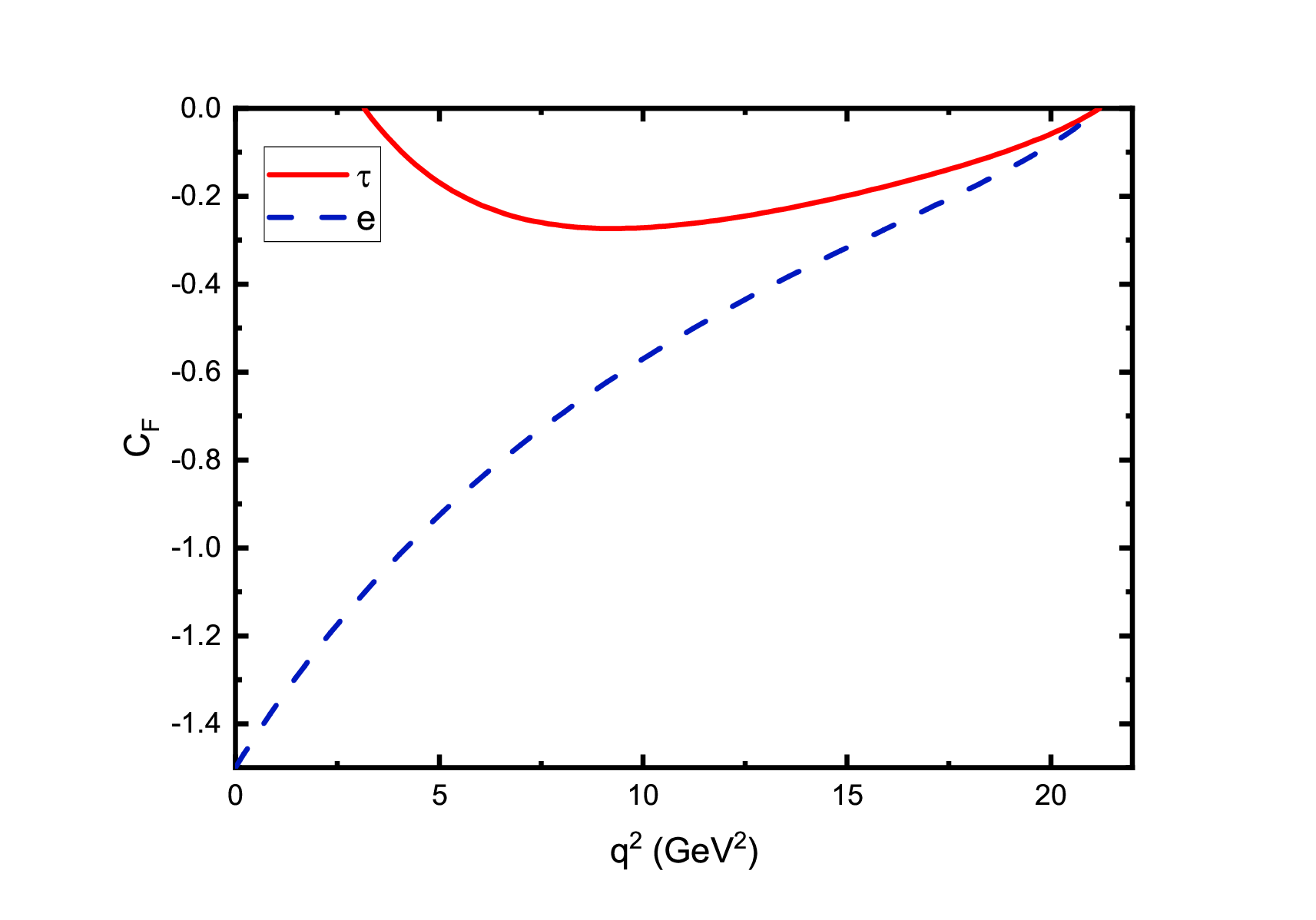}}
\hspace{-1.0cm}\subfigure{ \epsfxsize=8cm \epsffile{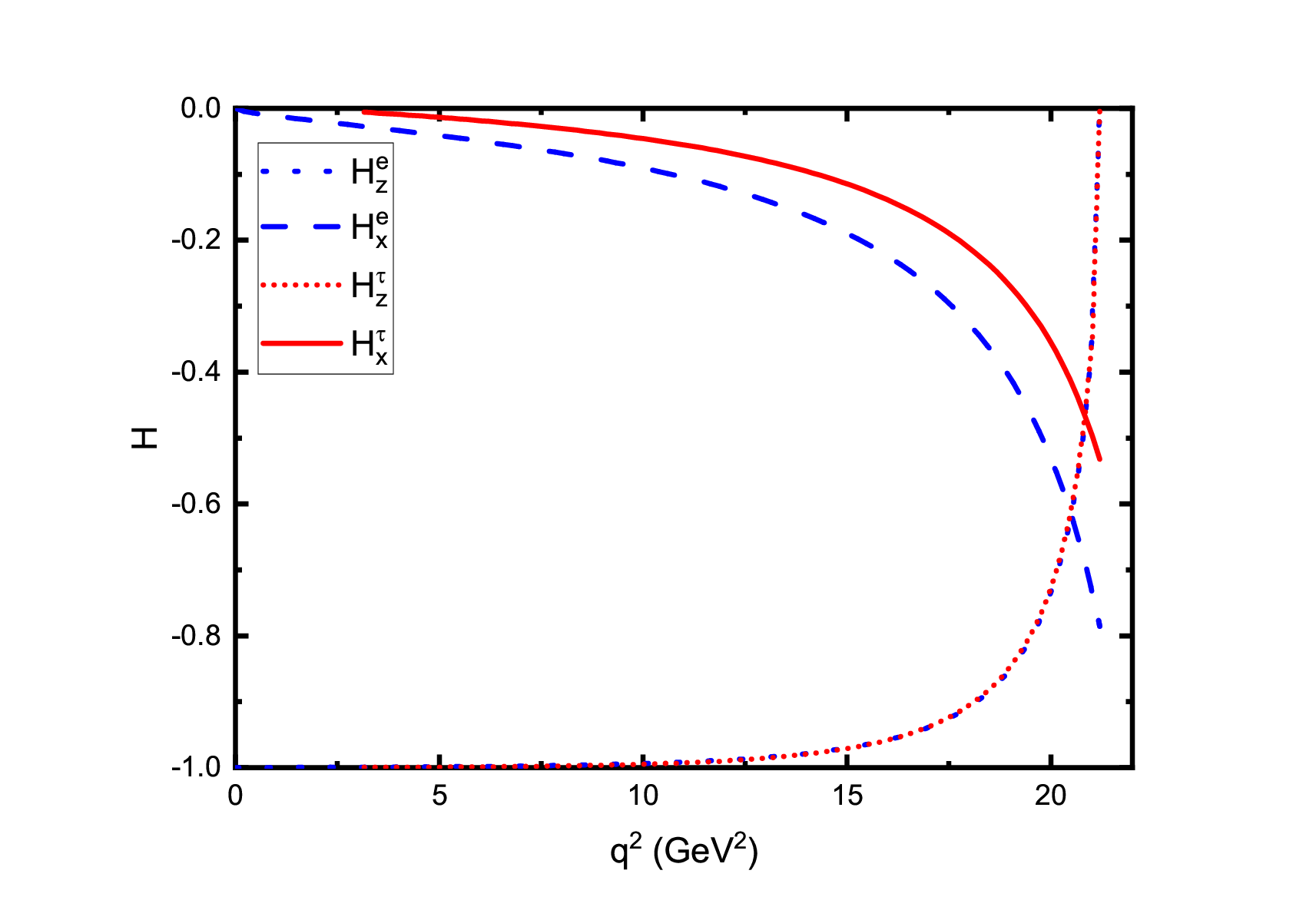}}}
\vspace{1cm}
\centerline{
\hspace{-1.0cm}\subfigure{ \epsfxsize=8cm \epsffile{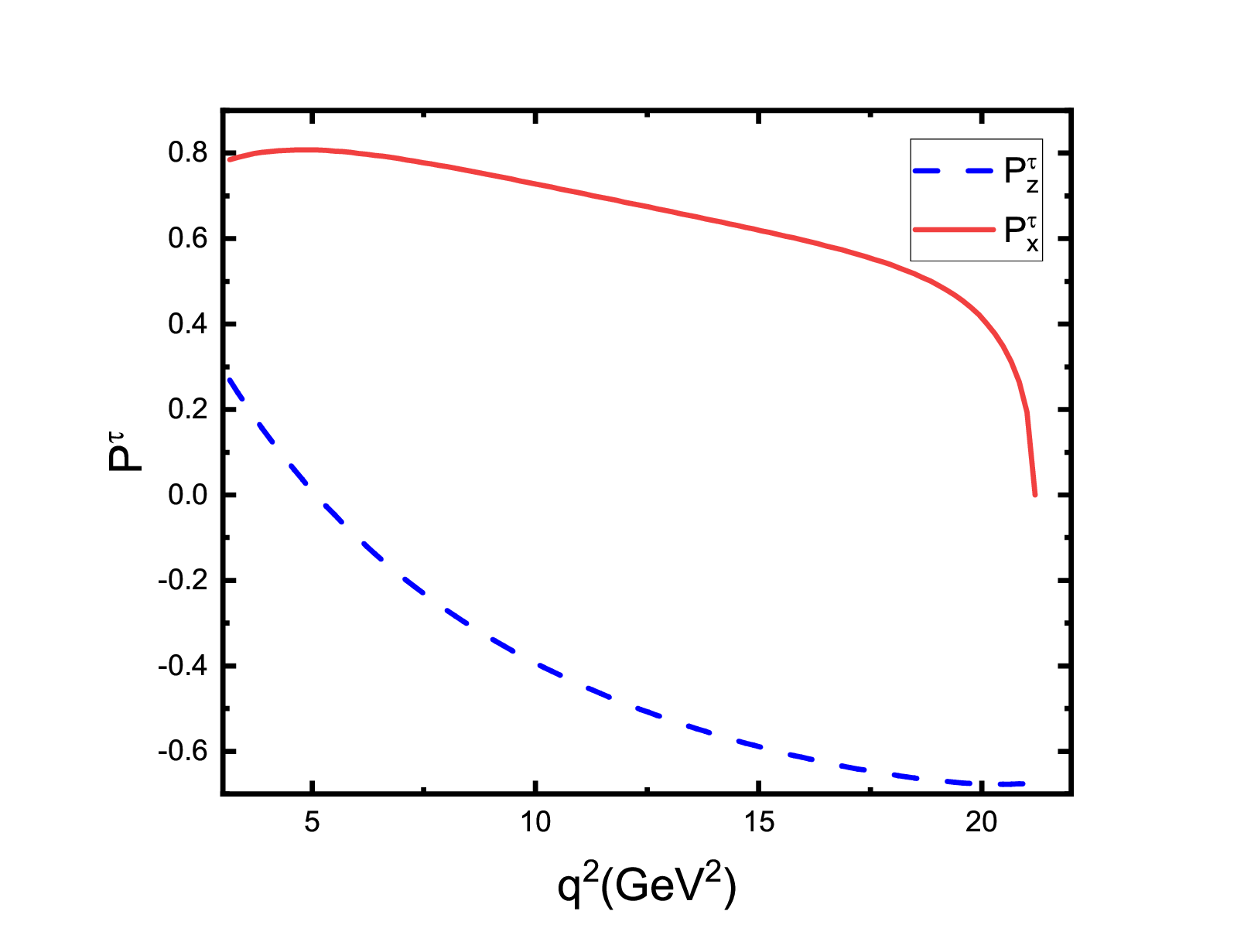}}
\hspace{-1.0cm}\subfigure{ \epsfxsize=8cm \epsffile{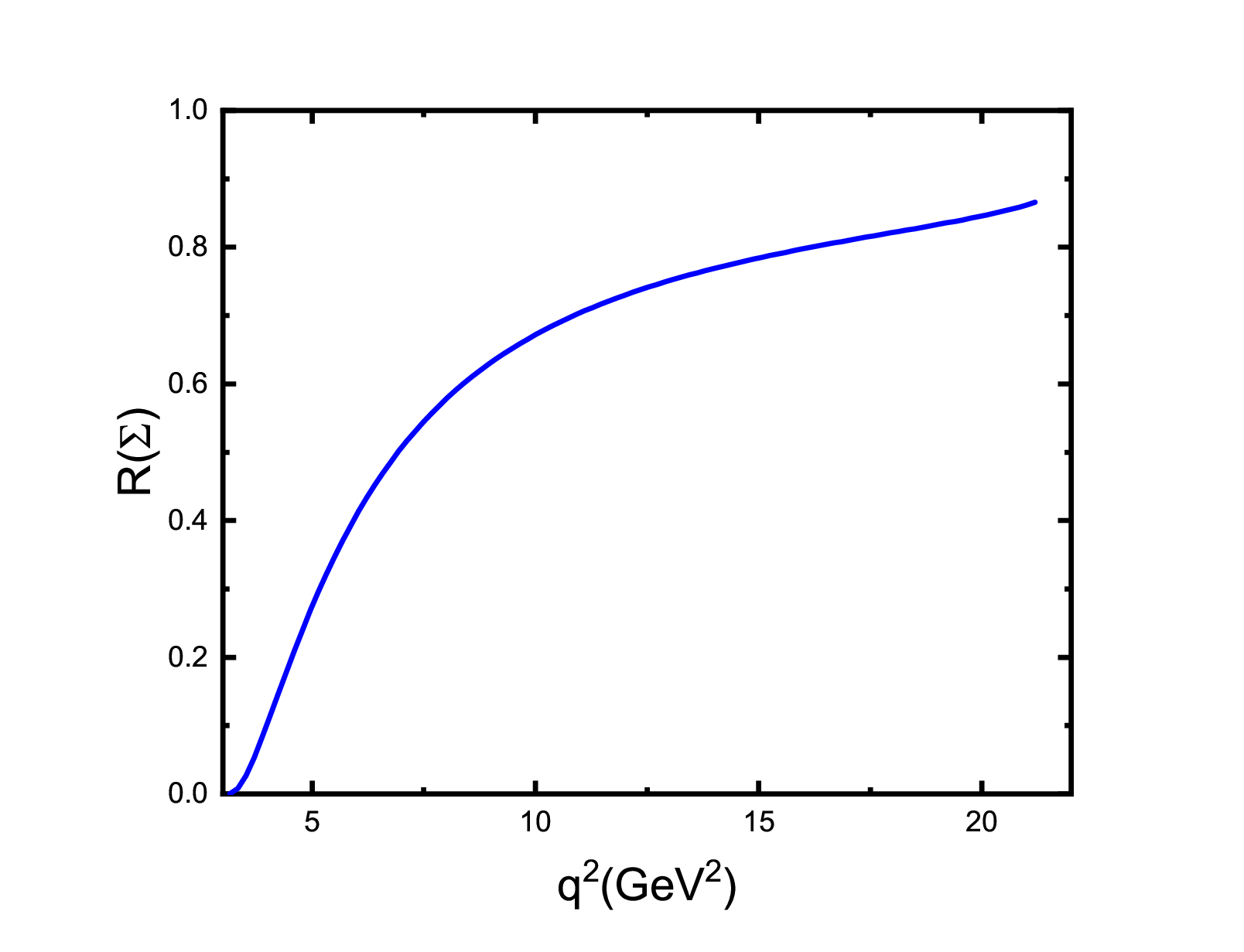}}}
\vspace{1cm}
\caption{The dependence of the observables  on $q^2$ for $\Xi_b^0\to \Sigma^+ \ell \nu_\ell$ decays.}
 \label{fig:q2p}
\end{center}
\end{figure}

\begin{figure}[!htbh]
\begin{center}
\setlength{\abovecaptionskip}{0pt}
\centerline{
\hspace{-1.0cm}\subfigure{ \epsfxsize=8cm \epsffile{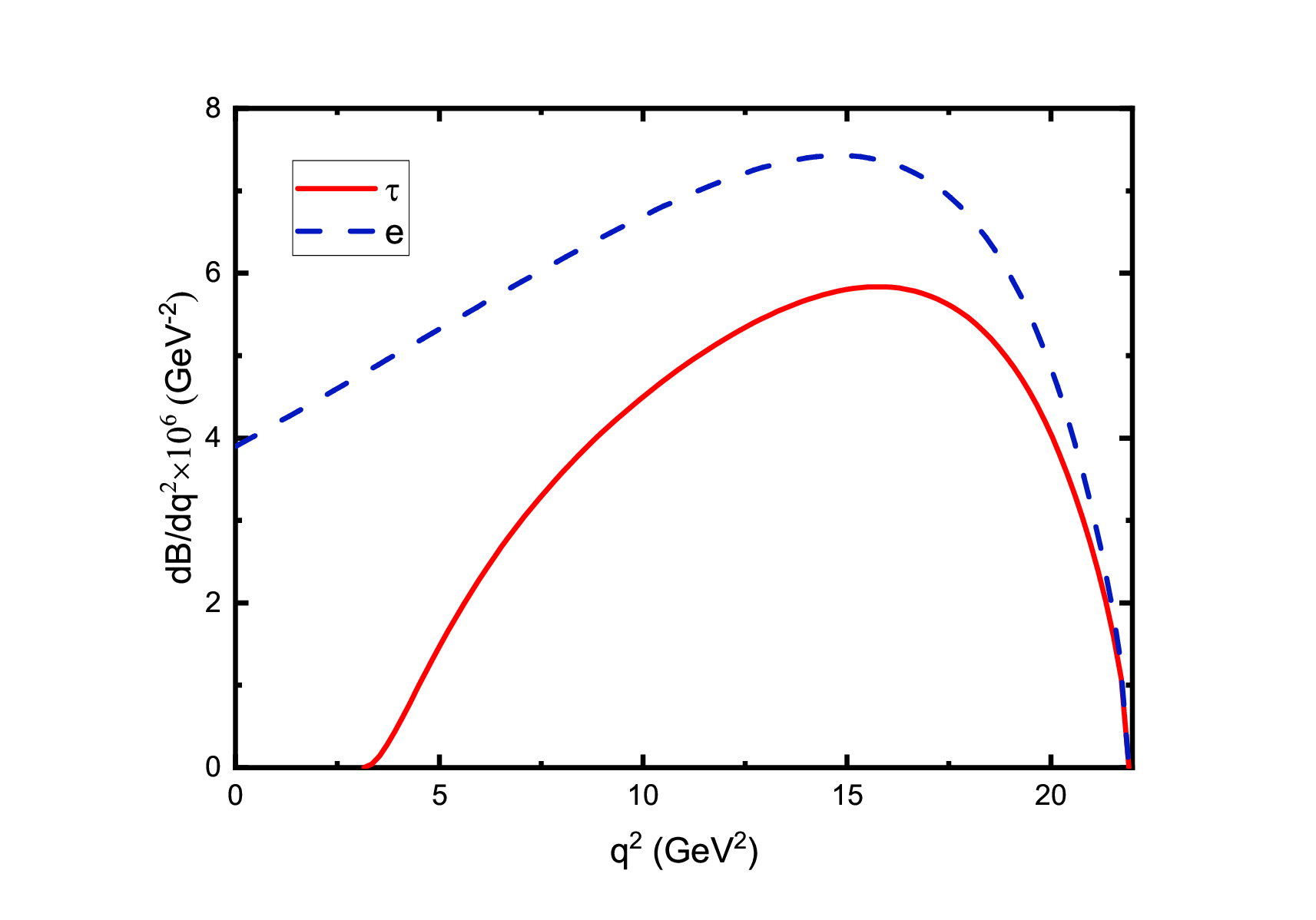}}
\hspace{-1.0cm}\subfigure{ \epsfxsize=8cm \epsffile{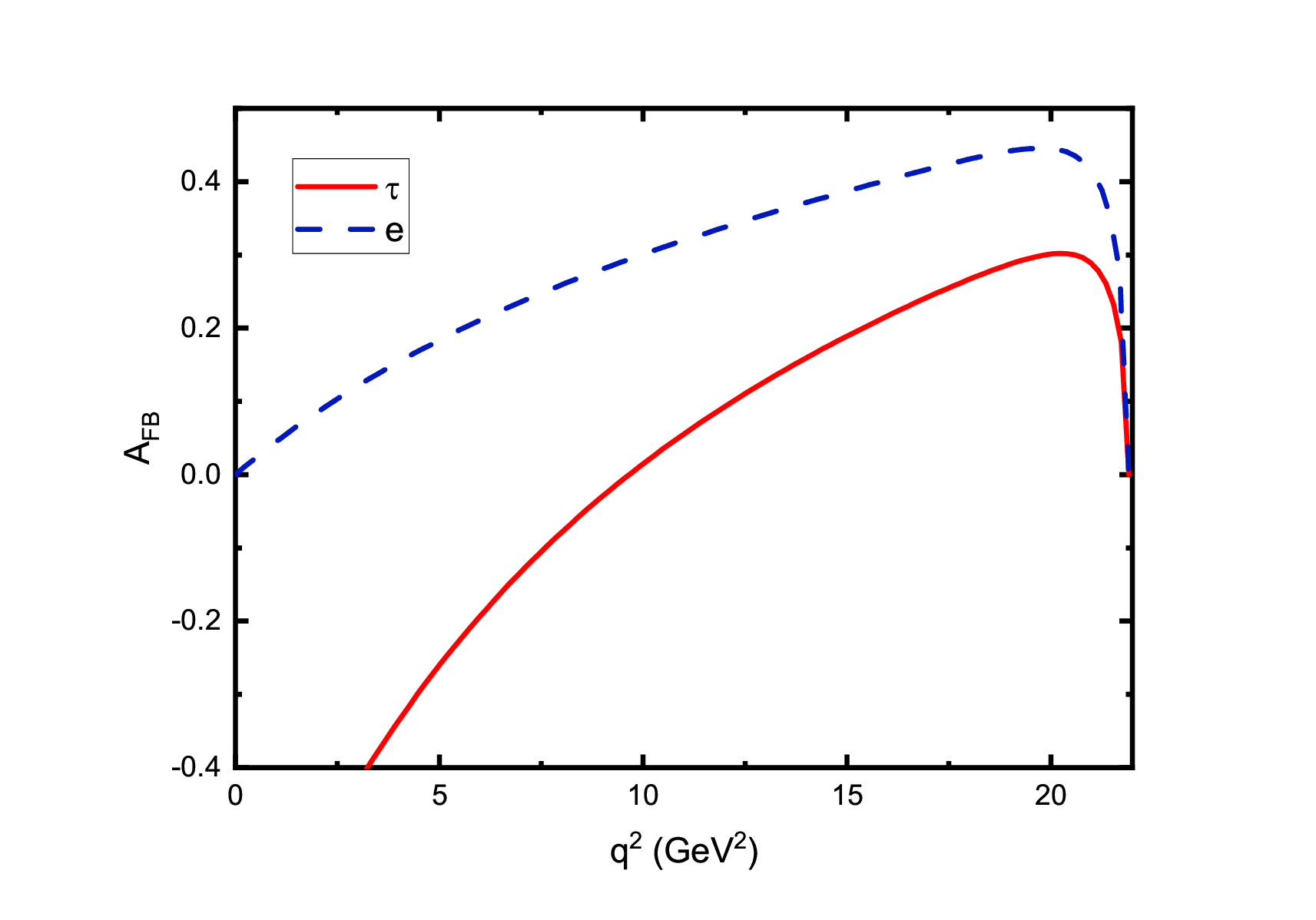}}}
\vspace{1cm}
\centerline{
\hspace{-1.0cm}\subfigure{ \epsfxsize=8cm \epsffile{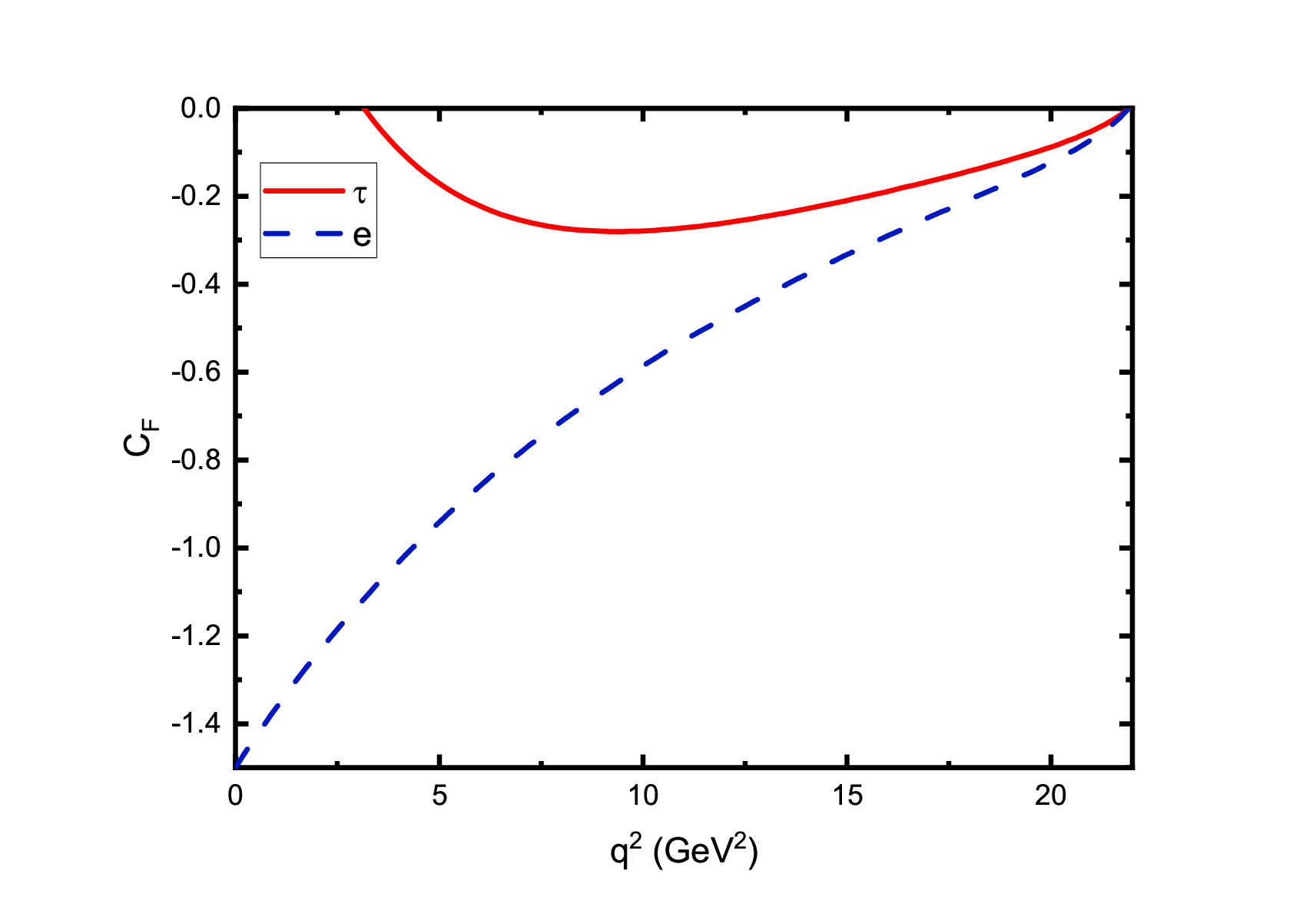}}
\hspace{-1.0cm}\subfigure{ \epsfxsize=8cm \epsffile{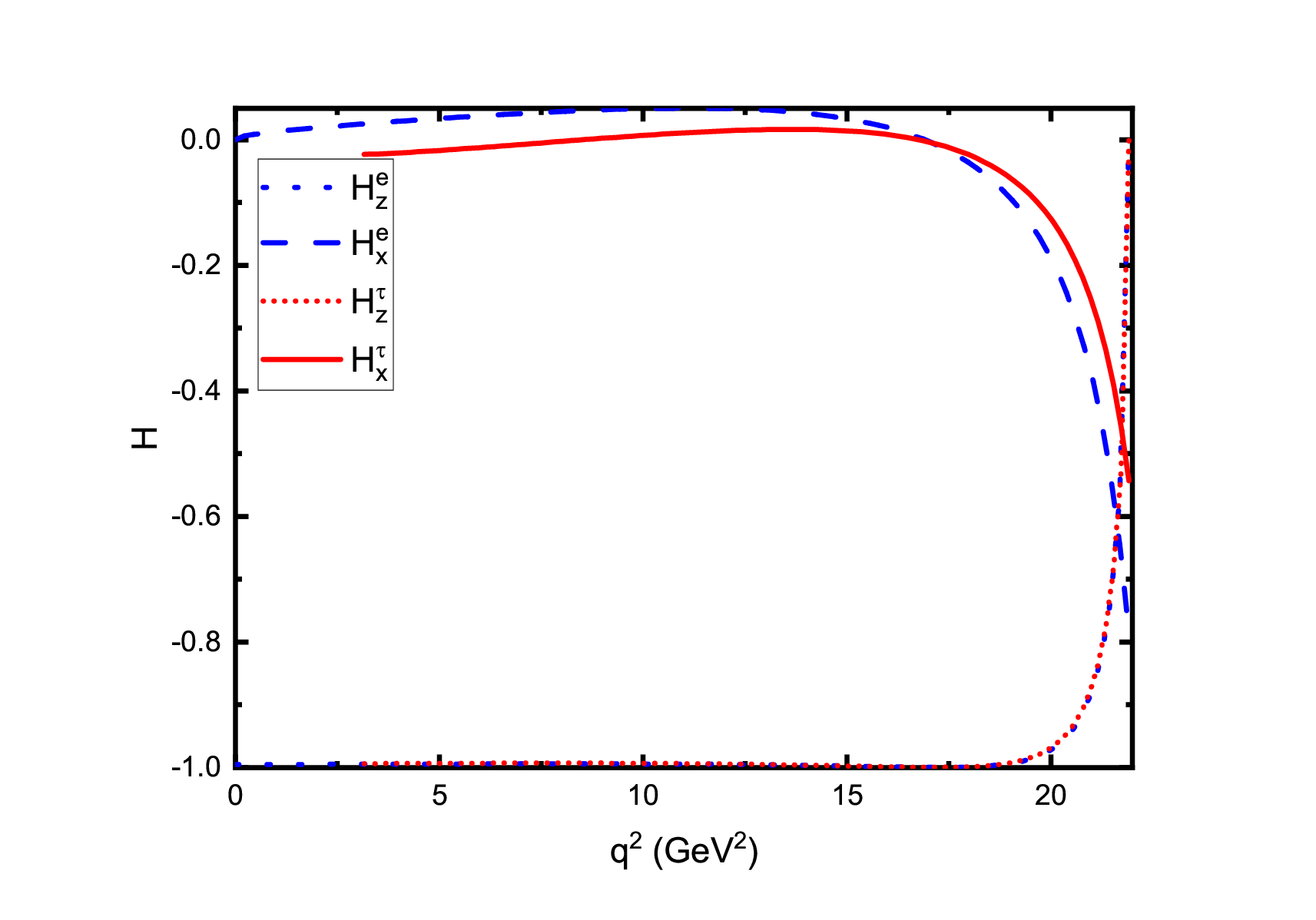}}}
\vspace{1cm}
\centerline{
\hspace{-1.0cm}\subfigure{ \epsfxsize=8cm \epsffile{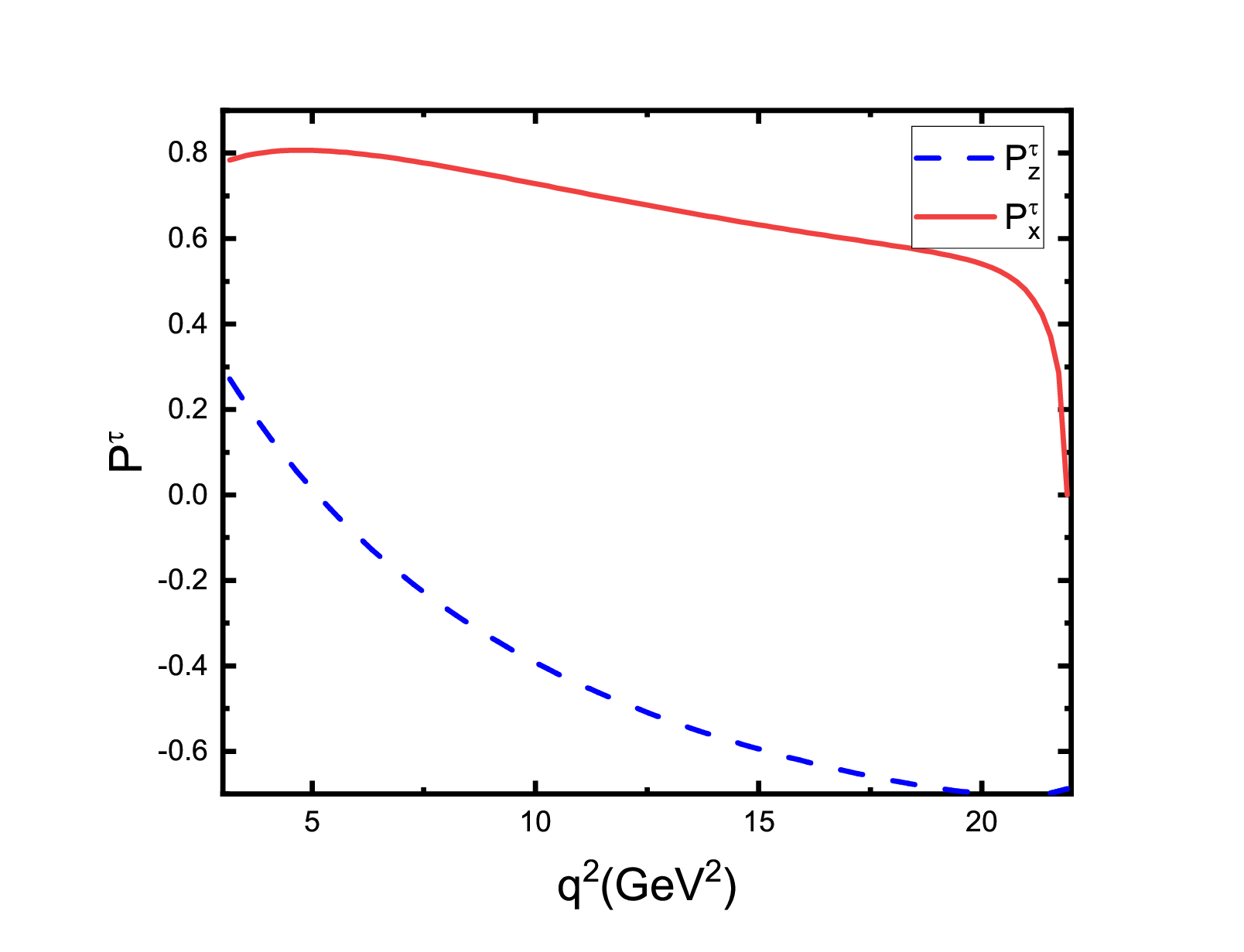}}
\hspace{-1.0cm}\subfigure{ \epsfxsize=8cm \epsffile{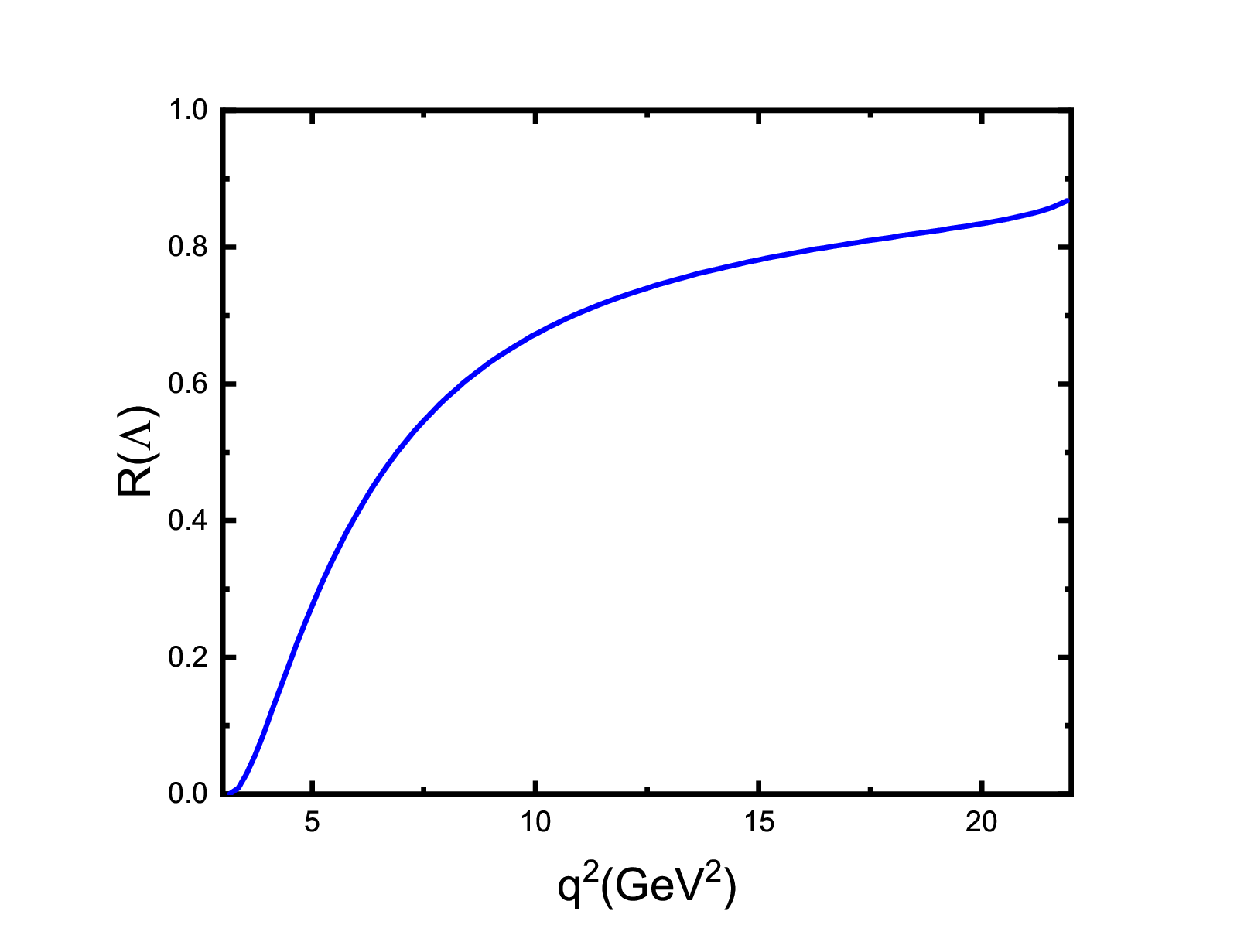}}}
\vspace{1cm}
\caption{Same as Fig.~\ref{fig:q2p} but for $\Xi_b^-\to \Lambda \ell \nu_\ell$ decays.}
 \label{fig:q2pp}
\end{center}
\end{figure}

Using the $q^2$ shapes of the form factors obtained in the $z$-series parametrization, we now can  analyze the  $q^2$ distributions of  the decay branching ratios and the angular observables. These distributions can disclose fascinating phenomena in specific $q^2$ regions, which may not be present in the integrated observables. As mentioned before, the asymmetries are less sensitive to the different LCDA models of $\Xi_b$, resulting in almost identical $q^2$ distributions that are difficult to distinguish. Hence we here take the Gegenbauer model as an example to show their $q^2$ dependence. The $q^2$ distributions of the differential branching ratios,  forward-backward asymmetry,  convexity parameter,  hardon polarizations,  lepton polarizations, and the LFU  ratio  are presented in Fig.~\ref{fig:q2p} for $\Xi_b^0\to \Sigma^+\ell \nu_\ell$ and Fig.~\ref{fig:q2pp} for $\Xi_b^-\to \Lambda \ell \nu_\ell$ decays. The $q^2$ dependence of  all of the observables are distinct for both $e$ and $\tau$ modes. Among these angular observables, the longitudinal hadronic polarization is independent of the lepton flavor; thus we observe that both  leptonic channels have almost identical $q^2$ distributions in the whole region, which are not distinguishable from the SM. This is an interesting observation to test in the experiment. Any significant shifts would exhibit large NP effects that appear through only tauonic interaction to the theory. In addition to this,  the $q^2$ behavior of all other observables for the $e$  mode is quite different from the $\tau$ mode. Since the weak interaction is purely left handed, the leptons are approximately absolutely polarized in the massless limit. Therefore, for the electron case, the longitudinal and transverse lepton polarizations are always constant values $-1$ and 0, respectively. We do not present their $q^2$ distributions here.

We now proceed to discuss the determination of $|V_{ub}/V_{cb}|$ from a combined analysis of the exclusive $b\to u \ell \nu_\ell$ and $b\to c \ell \nu_\ell$ tree-level processes. According to Eq.~(\ref{eq:gam}), the ratio of decay rates for the $b\to u \ell \nu_\ell$ and $b\to c \ell \nu_\ell$ processes can be written as
\begin{eqnarray}\label{eq:aaa}
\frac{\Gamma(b\to u \ell \nu_\ell)}{\Gamma(b\to c \ell \nu_\ell)}=\frac{|V_{ub}|^2}{|V_{cb}|^2}\mathcal{R}_\ell.
\end{eqnarray}
with
\begin{eqnarray}\label{eq:zzz}
\mathcal{R}_\ell=\frac{\int dq^2q^2|P|(1-\frac{m_\ell^2}{q^2})^2H(b\to u \ell \nu_\ell)}{\int dq^2q^2|P|(1-\frac{m_\ell^2}{q^2})^2H(b\to c \ell \nu_\ell)}.
\end{eqnarray}
In the $\Xi_b$ semileptonic decays, $H(b\to u \ell \nu_\ell)$ denotes the amplitude of the $\Xi_b^0\to \Sigma^+ \ell \nu_\ell$ ($\Xi_b^-\to \Lambda \ell \nu_\ell$)  semileptonic channel. Its partner in $b \to c \ell \nu_\ell$ transition corresponds to $H(\Xi_b^0\to \Xi_c^+ \ell \nu_\ell(\Xi_b^-\to \Xi_c^0 \ell \nu_\ell))$. $H(\Xi_b^0\to \Xi_c^+ \ell \nu_\ell)$  and $H(\Xi_b^-\to \Xi_c^0 \ell \nu_\ell)$ are associated with the isospin symmetry, whose analytical results have been derived in our preceding work~\cite{Rui:2025iwa}. Then we can calculate the ratio $\mathcal{R}_\ell$ in Eq.~(\ref{eq:zzz}) by integrating  the numerator and denominator with respect to $q^2$ separately.  The integration interval depends on the selected $q^2$ regions in forthcoming measurements. Once the experimental data of the decay-rate ratio in Eq.~(\ref{eq:aaa}) are available in the future, one can extract $|V_{ub}/V_{cb}|$ with theoretical calculations of $\mathcal{R}_\ell$. The PQCD predictions on the  $\mathcal{R}_\ell$ in the whole kinematical region with different LCDA models of $\Xi_b$ are given in Table~\ref{tab:ratio}. 
Since the results of $\Xi_b\to \Xi_c \ell \nu_\ell$ decays from the Free parton model are unavailable in~\cite{Rui:2025iwa}, we focus on the other three models for the numerical analysis.  
 Because the numerator and denominator in $\mathcal{R}_\ell$ share common sources of uncertainties, such as the input parameters from the $\Xi_b$ baryonic LCDAs and the hard scale $t$, the corresponding uncertainties are largely reduced in the ratio. The majority of uncertainties in $\mathcal{R}_\ell$ arise from the input parameters in the final state baryon LCDAs.
Since the parameters from different baryon LCDAs are not correlated, the resulting uncertainties have been combined in quadrature.
It is observed that the values of $\mathcal{R}_\ell$ in the last two rows are less sensitive to the three alternative models with a total uncertainty   around $10\%$. Therefore, we conclude that the combined analysis of the exclusive $\Xi_b^0\to \Sigma^+ \ell \nu_\ell$ and $\Xi_b^0\to \Xi_c^+ \ell \nu_\ell$  is more suitable for extracting $|V_{ub}/V_{cb}|$ in the $\Xi_b$ decays.
 Using the most precise results of $\mathcal{R}_\ell$ from the last column in Table~\ref{tab:ratio} and the latest average of $|V_{ub}/V_{cb}|=0.089\pm0.005$ from PDG~\cite{PDG2024}, one obtains the ratios of decay rates:
\begin{eqnarray}\label{eq:rrr}
\frac{\Gamma(\Xi_b^-\to \Lambda \tau \nu_\tau)}{\Gamma(\Xi_b^-\to \Xi_c^0 \tau \nu_\tau)}&=&(2.2\pm 0.4\pm0.2)\times 10^{-3},   \nonumber\\
\frac{\Gamma(\Xi_b^-\to \Lambda e \nu_e)}{\Gamma(\Xi_b^-\to \Xi_c^0 e \nu_e)}&=&(1.2\pm 0.2\pm0.1)\times 10^{-3},  \nonumber\\
\frac{\Gamma(\Xi_b^0\to \Sigma^+ \tau \nu_\tau)}{\Gamma(\Xi_b^0\to \Xi_c^+ \tau \nu_\tau)}&=&(1.4\pm 0.1\pm 0.2)\times 10^{-2}, \nonumber\\ 
\frac{\Gamma(\Xi_b^0\to \Sigma^+ e \nu_e)}{\Gamma(\Xi_b^0\to \Xi_c^+ e \nu_e)}&=&(6.7^{+0.5}_{-0.6}\pm0.8)\times 10^{-3},
\end{eqnarray}
where the first uncertainty originates from the variations of the theory parameters and the second from the experimental input. These ratios can be checked experimentally.
\begin{table}[!htbh]
	\caption{PQCD predictions on the $\mathcal{R}_\ell$ in the full $q^2$ region with different LCDA models of $\Xi_b$. }
	\label{tab:ratio}
	\begin{tabular}[t]{lccc}
	\hline\hline
$\mathcal{R}_\ell$    & S1  &S2  &S3 \\ \hline
$\mathcal{R}_\tau(\Lambda/\Xi_c)$     &  $0.23^{+0.10}_{-0.08}$             &  $0.22^{+0.09}_{-0.07}$        &  $0.28^{+0.05}_{-0.05}$ \\ 
$\mathcal{R}_e(\Lambda/\Xi_c)$        &  $0.113^{+0.045}_{-0.034}$             &  $0.106^{+0.042}_{-0.032}$      &  $0.147^{+0.025}_{-0.031}$  \\ 
$\mathcal{R}_\tau(\Sigma/\Xi_c)$     &  $2.1\pm 0.2$             &  $2.2\pm 0.2$      &  $1.8^{+0.1}_{-0.2}$   \\ 
$\mathcal{R}_e(\Sigma/\Xi_c)$        &  $0.94\pm 0.10$             &  $0.99\pm 0.07$      &  $0.85^{+0.06}_{-0.08}$\\ 
\hline\hline
	\end{tabular}
\end{table}

\section{Summary}\label{sec:sum} 

In this work, we present a detailed study of the semileptonic decays $\Xi_b\to (\Lambda,\Sigma) \ell \nu_\ell$ within the framework of PQCD. To address the present uncertainty in the $\Xi_b$ baryon's LCDAs, we comparatively evaluate four models: the Exponential, QCDSR, Gegenbauer, and Free Parton models. The corresponding form factors are computed at the maximum recoil point ($q^2=0$). The results from all models are mutually consistent within uncertainties and agree with predictions from the heavy quark limit. We find that predictions from the Exponential and Free Parton models carry significant theoretical uncertainties, which are dominated by their model parameters. In contrast, the QCDSR and Gegenbauer models show less parameter sensitivity and are consequently more stable. Where available, our numerical results are compared with previous calculations. The form factors are extended over the full physical region using a $z$-series parametrization, which effectively describes their $q^2$ evolution and enables a smooth extrapolation. All form factors grow in magnitude with increasing $q^2$, as anticipated. Among them, $f_1(q^2)$ and $g_1(q^2)$ display comparable $q^2$ dependence and constitute the dominant contributions. In contrast, the form factors $f_3(q^2)$ and $g_3(q^2)$ are found to be opposite in sign to the others for all $q^2$.
 
With the form factors determined, we proceed to predict the differential and integrated angular observables, decay branching ratios, and LFU ratios. The branching fractions for the $\Xi_b\to \Sigma \ell \nu_\ell$ and $\Xi_b\to \Lambda \ell \nu_\ell$ channels are predicted to be of order $10^{-4}$ and $10^{-5}$, respectively, making them accessible to measurement in the near future by the LHCb experiment. The dominant theoretical uncertainty originates from the nonperturbative parameters of the baryonic LCDAs. Consequently, more precise determinations of these baryonic LCDAs are required to improve the accuracy of our calculations. We also predict the LFU ratios, finding $\mathcal{R}_{\Sigma(\Lambda)}\sim 0.6$. These values are compatible with the SM expectation as exemplified by the corresponding ratio $\mathcal{R}_{\pi}$ in $B\to \pi \ell \nu_\ell$ decays, and they await confrontation with future experimental data. Furthermore, the calculated angular observables demonstrate good stability against variations in model parameters and across different LCDA models, though lepton-mass effects are found to be significant for their average values.

Finally, we investigate the ratios $\mathcal{R}_\ell(\Lambda/\Xi_c)$ and $\mathcal{R}_\ell(\Sigma/\Xi_c)$ over the entire kinematical range. The latter is calculated with high precision and exhibits minimal sensitivity to the choice of LCDA model. A future measurement of the ratio $\Gamma(\Xi_b^0\to \Sigma^+ \ell \nu_\ell) / \Gamma(\Xi_b^0\to \Xi_c^+ \ell \nu_\ell)$ would provide a fully complementary avenue for extracting the CKM matrix element ratio $|V_{ub}/V_{cb}|$ from baryonic decays. Our comprehensive findings provide crucial theoretical input for probing potential new physics effects beyond the SM in these decay channels.

\begin{acknowledgments}
This work is supported by the National Natural Science Foundation of China under Grants No. 12075086, No.12375089 and No.12435004. Z.T.Z. is  supported by  the Natural Science Foundation of Shandong Province under  Grant No. ZR2022MA035. Y.Li is also supported by the Natural Science Foundation of Shandong Province under  Grant No. ZR2022ZD26.
\end{acknowledgments}

\begin{appendix}
\section{LIGHT-CONE DISTRIBUTION AMPLITUDES}\label{sec:LCDAs}

Similar to $\Lambda_b$ baryon LCDAs~\cite{plb665197,jhep112013191,epjc732302,plb738334,jhep022016179,Ali:2012zza}, the LCDAs of $\Xi_b$ baryon up to twist-4 accuracy in the momentum space can be written as~\cite{Rui:2025iwa,Rui:2023fiz}
\begin{eqnarray}\label{eq:LCDAs20}
(\Psi_{\mathcal{B}_i})_{\alpha\beta\gamma}(x_i) &=&\frac{1}{8N_c}\left\{\frac{ f^{(1)}}{\sqrt{2}}
[(\slashed{ n}\gamma_5C)_{\alpha\beta}\phi_2(x_2,x_3)+(\slashed{ v}\gamma_5C)_{\alpha\beta}\phi_4(x_2,x_3)](u_{\mathcal{B}_i})_\gamma \right.\nonumber\\
&& \left.+f^{(2)}[(\gamma_5C)_{\alpha\beta}\phi_{3s}(x_2,x_3)-\frac{i}{2}(\sigma_{ nv}\gamma_5C)_{\alpha\beta}\phi_{3a}(x_2,x_3)](u_{\mathcal{B}_i})_\gamma\right\},
\end{eqnarray}
where two light-cone vectors $n=(1,0,\textbf{0}_T)$ and $v=(0,1,\textbf{0}_T)$ satisfy $n\cdot v=1$. $N_c$ is the number of colors. $C$ denotes the charge conjugation matrix. The decay constants for the $\Xi_b$ baryon take the values   $f^{(1,2)}=(0.032\pm 0.009)$ GeV$^3$ from the QCD sum rules~\cite{Wang:2010fq}. The functions $\phi_{ 2,4,3s,3a}$   represent four different LCDAs, and the numbers in the subscript indicate the twist. Their asymptotic forms with four popular models are summarized in~\cite{Rui:2025iwa}.

For an outgoing light baryon, the corresponding LCDAs up to twist six in the  momentum space is given below~\cite{Braun:2000kw}:
 \begin{eqnarray}\label{eq:LCDAsp}
 (\bar{\Psi}_{\mathcal{B}_f})_{\alpha\beta\gamma}(x'_i) &=&  -\frac{1}{8\sqrt{2}N_c} [ \mathcal{S}_1m( \bar{u}_{\mathcal{B}_f}\gamma_5)_\gamma C_{\beta\alpha}+
\mathcal{S}_2m^2( \bar{u}_{\mathcal{B}_f}\gamma_5\rlap{/}{z})_\gamma C_{\beta\alpha}+  \mathcal{P}_1m( \bar{u}_{\mathcal{B}_f})_\gamma (C\gamma_5)_{\beta\alpha}   \nonumber\\&&
+ \mathcal{P}_2m^2( \bar{u}_{\mathcal{B}_f}\rlap{/}{z})_\gamma (C\gamma_5)_{\beta\alpha} +\mathcal{V}_1( \bar{u}_{\mathcal{B}_f}\gamma_5)_\gamma (C\rlap{/}{p'})_{\beta\alpha}
+\mathcal{V}_2m( \bar{u}_{\mathcal{B}_f}\gamma_5\rlap{/}{z})_\gamma (C\rlap{/}{p'})_{\beta\alpha} \nonumber\\&& +\mathcal{V}_3m( \bar{u}_{\mathcal{B}_f}\gamma_5 \gamma_\mu)_\gamma (C\gamma^\mu)_{\beta\alpha}+\mathcal{V}_4m^2( \bar{u}_{\mathcal{B}_f}\gamma_5)_\gamma (C\rlap{/}{z})_{\beta\alpha}
-i\mathcal{V}_5m^2(  \bar{u}_{\mathcal{B}_f}\gamma_5\sigma^{\mu\nu}z_{\nu})_\gamma( C\gamma_\mu)_{\beta\alpha} \nonumber\\&&
+\mathcal{V}_6m^3( \bar{u}_{\mathcal{B}_f}\gamma_5\rlap{/}{z})_\gamma (C\rlap{/}{z})_{\beta\alpha}  +\mathcal{A}_1( \bar{u}_{\mathcal{B}_f})_\gamma (C\gamma_5\rlap{/}{p'})_{\beta\alpha}
+\mathcal{A}_2m( \bar{u}_{\mathcal{B}_f}\rlap{/}{z})_\gamma (C\gamma_5\rlap{/}{p'})_{\beta\alpha} \nonumber\\&&+\mathcal{A}_3m( \bar{u}_{\mathcal{B}_f}\gamma_\mu)_\gamma (C\gamma_5\gamma^\mu)_{\beta\alpha}+\mathcal{A}_4m^2( \bar{u}_{\mathcal{B}_f})_\gamma (C\gamma_5\rlap{/}{z})_{\beta\alpha}
-i\mathcal{A}_5m^2(  \bar{u}_{\mathcal{B}_f}\sigma^{\mu\nu}z_{\nu})_\gamma( C\gamma_5\gamma_\mu)_{\beta\alpha} \nonumber\\&&
+\mathcal{A}_6m^3( \bar{u}_{\mathcal{B}_f}\rlap{/}{z})_\gamma (C\gamma_5\rlap{/}{z})_{\beta\alpha}
-i\mathcal{T}_1( \bar{u}_{\mathcal{B}_f} \gamma_5\gamma^{\mu})_\gamma( C \sigma_{\mu\nu}p'^\nu)_{\beta\alpha}
-i\mathcal{T}_2m( \bar{u}_{\mathcal{B}_f} \gamma_5)_\gamma( C \sigma_{\mu\nu}p'^\mu z^\nu)_{\beta\alpha} \nonumber\\&&
+\mathcal{T}_3m( \bar{u}_{\mathcal{B}_f} \gamma_5 \sigma^{\mu\nu} )_\gamma( C \sigma_{\mu\nu})_{\beta\alpha}
+\mathcal{T}_4m( \bar{u}_{\mathcal{B}_f} \gamma_5 \sigma^{\mu\rho}z_\rho )_\gamma( C \sigma_{\mu\nu}p'^\nu)_{\beta\alpha}\nonumber\\&&
-i\mathcal{T}_5m^2( \bar{u}_{\mathcal{B}_f} \gamma_5\gamma^{\mu})_\gamma( C \sigma_{\mu\nu}z^\nu)_{\beta\alpha}
-i\mathcal{T}_6m^2( \bar{u}_{\mathcal{B}_f} \gamma_5\rlap{/}{z})_\gamma( C \sigma_{\mu\nu}z^\mu p'^\nu)_{\beta\alpha} \nonumber\\&&
+\mathcal{T}_7m^2( \bar{u}_{\mathcal{B}_f} \gamma_5\rlap{/}{z} \sigma^{\mu\rho} )_\gamma( C \sigma_{\mu\nu})_{\beta\alpha}
+\mathcal{T}_8m^3( \bar{u}_{\mathcal{B}_f} \gamma_5 \sigma^{\mu\rho} z_\rho)_\gamma( C \sigma_{\mu\nu}z^\nu)_{\beta\alpha}],
\end{eqnarray}
where $z$ is an arbitrary light like vector with $z^2=0$. We choose $z=\frac{\sqrt{2}}{Mf^+}v$ so that $z\cdot p'=1$~\cite{Braun:1999te}. To arrive at the above expression,  we performed a transpose conjugate transformation on a baryon to vacuum matrix element. Note that the ``calligraphic" invariant functions  $\mathcal{S}, \mathcal{P}, \mathcal{A}, \mathcal{V}, \mathcal{T}$ in Eq.~(\ref{eq:LCDAsp}) do not have a definite twist. They can be expanded in terms of 24 LCDAs with definite twist and symmetry, which depend on the longitudinal momentum fractions carried by the quarks inside the baryon. These relations and explicit expressions of various twist LCDAs  can be found in~\cite{Braun:1999te,Liu:2008yg} and will not be duplicated here.

\section{FACTORIZATION FORMULAS}\label{sec:for}

\begin{table}[!htbh]
\caption{Internal gluon virtualities $t_{A,B}$ and quark virtualities $t_{C,D}$ for each diagram in Eq.~(\ref{eq:ttt}). }
\label{tab:ttt}
\begin{tabular}[t]{lcccc}
\hline\hline
 $\xi$      & $t_A$ & $t_B$ & $t_C$  & $t_D$                 \\ \hline
$a$    & $x_3x_3'f^+M^2$  & $(1-x_1)(1-x_1')f^+M^2$  & $(1-x_1)x'_3f^+M^2$   & $(1-x_1')f^+M^2$                          \\
$b$    & $x_3x_3'f^+M^2$  & $(1-x_1)(1-x_1')f^+M^2$  & $(1-x'_1)x_3f^+M^2$   & $(1-x_1')f^+M^2$                           \\
$c$    & $x_3x_3'f^+M^2$      & $x_2x_2'f^+M^2$    & $(x_2+(1-x_2)x'_3f^+)M^2$   & $(1-x_1')f^+M^2$                          \\
$d$    & $x_3x_3'f^+M^2$      & $x_2x_2'f^+M^2$     & $x_3(1-x'_2)f^+M^2$   & $(x_3+x'_2(1-x_3)f^+)M^2$                          \\
$e$    & $x_3x_3'f^+M^2$      & $(1-x_1)(1-x_1')f^+M^2$    &  $(1-x_1)x'_3f^+M^2$    & $(1-x_1)f^+M^2$                           \\
$f$    & $x_3x_3'f^+M^2$      & $(1-x_1)(1-x_1')f^+M^2$  &  $(1-x'_1)x_3f^+M^2$   & $(1-x_1)f^+M^2$                           \\
$g$    & $x_3x_3'f^+M^2$   & $x_2x_2'f^+M^2$     &  $(1-x_1)f^+M^2$ &   $(1-x'_3)x_2f^+M^2$                    \\

$h$    & $(1-x_1)(1-x_1')f^+M^2$ & $x_2x_2'f^+M^2$  & $(1-x'_1)f^+M^2$   & $(1-x_1)x'_2f^+M^2$                          \\
$i$    & $(1-x_1)(1-x_1')f^+M^2$ & $x_2x_2'f^+M^2$  & $(1-x_1')f^+M^2$  & $(1-x'_1)x_2f^+M^2$  \\
$j$    & $x_3x_3'f^+M^2$      & $x_2x_2'f^+M^2$     & $(1-x_1')f^+M^2$   & $(x_3+x'_2(1-x_3)f^+)M^2$                          \\
$k$    & $x_3x_3'f^+M^2$      & $x_2x_2'f^+M^2$     & $(x_2+x'_3(1-x_2)f^+)M^2$ & $x_2(1-x'_3)f^+M^2$   \\
$l$    & $(1-x_1)(1-x_1')f^+M^2$& $x_2x_2'f^+M^2$    & $(1-x_1)f^+M^2$  &  $(1-x_1)x'_2f^+M^2$    \\
$m$    & $(1-x_1)(1-x_1')f^+M^2$& $x_2x_2'f^+M^2$   & $(1-x_1)f^+M^2$ &  $(1-x'_1)x_2f^+M^2$   \\
$n$    & $x_3x_3'f^+M^2$   & $x_2x_2'f^+M^2$        &   $(1-x'_2)x_3f^+M^2$ &  $(1-x_1)f^+M^2$ \\
\hline\hline
\end{tabular}
\end{table}
From Fig.~\ref{fig:C}, there are four virtualities of internal particles in each diagram, two of which correspond to hard gluons ($t_{A,B}$) and two to virtual quarks ($t_{C,D}$). The hard scale $t_\xi$ in Eq.~(\ref{eq:FG}) is chosen as the maximal value among them, including the factorization scales in a hard amplitude,
\begin{eqnarray}\label{eq:ttt}
t_\xi=\max(\sqrt{|t_A|},\sqrt{|t_B|},\sqrt{|t_C|},\sqrt{|t_D|},w,w'),
\end{eqnarray}
where  $w$ and  $w^{'}$ are the factorization scales in the Sudakov resummations associated with the initial and final states. The hard scales $t_\xi$ for each diagram are collected in Table~\ref{tab:ttt}. Table~\ref{tab:bb} present the expressions of $\Omega_\xi $, where the Bessel functions $K_{0}$ and the auxiliary functions $h_{1,2}$ can be found in~\cite{Rui:2025iwa}. The formulas of $H_\xi(x_i,x'_i)$ are rather lengthy due to many higher-twist contributions being included. Here, we only show some details for the first diagram in Fig~\ref{fig:C}. The others can be derived in a similar way. The vector form factors $ H^{F_i}_a(x_i,x'_i)$ are as below
\begin{eqnarray}
H^{F_1}_a(x_i,x'_i)&=&-\frac{r^2 M^3}{r-1}(x_2+x_3) \phi _2 (2 \mathcal{A}_3-4 \mathcal{A}_5-\mathcal{P}_1-2 \mathcal{P}_2+\mathcal{S}_1-2
   \mathcal{S}_2+\mathcal{T}_1-\mathcal{T}_2-\mathcal{T}_4\nonumber\\&&+2 \mathcal{T}_5-2 \mathcal{T}_6+2 \mathcal{V}_3+4 \mathcal{V}_5+r \mathcal{A}_1 x'_1+2 r
   \mathcal{A}_2 x'_1+2 r \mathcal{A}_4 x'_1+2 r \mathcal{A}_5 x'_1+4 r \mathcal{A}_6 x'_1-2 r \mathcal{T}_1 x'_1\nonumber\\&&+4 r \mathcal{T}_4 x'_1-4 r \mathcal{T}_5
   x'_1+8 r \mathcal{T}_8 x'_1-r \mathcal{V}_1 x'_1+2 r \mathcal{V}_2 x'_1-2 r \mathcal{V}_4 x'_1-2 r \mathcal{V}_5 x'_1+4 r \mathcal{V}_6 x'_1)
\nonumber\\&&+\frac{\phi _4 x'_3 M^3}{r-1}(-\mathcal{A}_1-2 r \mathcal{A}_3+r \mathcal{P}_1-r \mathcal{S}_1-r \mathcal{T}_1+2 \mathcal{T}_1-r \mathcal{T}_2+r
   \mathcal{T}_4+\mathcal{V}_1-2 r \mathcal{V}_3)\nonumber\\&&-\frac{2 r \phi _{3 s} M^3}{r-1}(2 x_2 \mathcal{A}_3+2 x_3 \mathcal{A}_3-2 r x_2
   x'_1 \mathcal{A}_3-2 r x_3 x'_1 \mathcal{A}_3+2 r x'_3 \mathcal{A}_3-2 x'_3 \mathcal{A}_3+x_2 \mathcal{P}_1\nonumber\\&&+x_3 \mathcal{P}_1+x_2 \mathcal{S}_1+x_3
   \mathcal{S}_1-x_2 \mathcal{T}_1-x_3 \mathcal{T}_1-x_2 \mathcal{T}_2-x_3 \mathcal{T}_2+4 x_2 \mathcal{T}_3+4 x_3 \mathcal{T}_3+x_2 \mathcal{T}_4\nonumber\\&&+x_3
   \mathcal{T}_4-2 x_2 \mathcal{V}_3-2 x_3 \mathcal{V}_3-r x_2 \mathcal{A}_1 x'_1-r x_3 \mathcal{A}_1 x'_1-2 r x_2 \mathcal{A}_4 x'_1-2 r x_3 \mathcal{A}_4\nonumber\\&&
   x'_1+2 r x_2 \mathcal{A}_5 x'_1+2 r x_3 \mathcal{A}_5 x'_1+2 r x_2 \mathcal{T}_1 x'_1+2 r x_3 \mathcal{T}_1 x'_1-8 r x_2 \mathcal{T}_3 x'_1-8 r x_3
   \mathcal{T}_3 x'_1\nonumber\\&&+4 r x_2 \mathcal{T}_5 x'_1+4 r x_3 \mathcal{T}_5 x'_1+16 r x_2 \mathcal{T}_7 x'_1+16 r x_3 \mathcal{T}_7 x'_1-r x_2 \mathcal{V}_1
   x'_1\nonumber\\&&-r x_3 \mathcal{V}_1 x'_1+2 r x_2 \mathcal{V}_3 x'_1+2 r x_3 \mathcal{V}_3 x'_1-2 r x_2 \mathcal{V}_4 x'_1-2 r x_3 \mathcal{V}_4 x'_1+2 r x_2
   \mathcal{V}_5 x'_1\nonumber\\&&+2 r x_3 \mathcal{V}_5 x'_1-\mathcal{A}_1 x'_3-2 \mathcal{A}_2 x'_3-4 r \mathcal{A}_5 x'_3+r \mathcal{P}_1 x'_3+2 r \mathcal{P}_2 x'_3+r
   \mathcal{S}_1 x'_3\nonumber\\&&-2 r \mathcal{S}_2 x'_3-r \mathcal{T}_1 x'_3+2 \mathcal{T}_1 x'_3+r \mathcal{T}_2 x'_3+4 r \mathcal{T}_3 x'_3-8 \mathcal{T}_3 x'_3+r
   \mathcal{T}_4 x'_3\nonumber\\&&-4 \mathcal{T}_4 x'_3-2 r \mathcal{T}_5 x'_3+2 r \mathcal{T}_6 x'_3-8 r \mathcal{T}_7 x'_3-\mathcal{V}_1 x'_3+2 \mathcal{V}_2 x'_3-2 r
   \mathcal{V}_3 x'_3+2 \mathcal{V}_3 x'_3-4 r \mathcal{V}_5 x'_3),
\end{eqnarray}

\begin{eqnarray}
H^{F_2}_a(x_i,x'_i)&=&-\frac{2 r^3 M^3}{(r-1) (r+1)^2} (x_2+x_3) \phi _2 (-2 r x'_1 \mathcal{A}_4-2 \mathcal{A}_4+2 \mathcal{A}_5-4 \mathcal{A}_6+\mathcal{P}_1+2
   \mathcal{P}_2-\mathcal{S}_1+2 \mathcal{S}_2+\mathcal{T}_1+\mathcal{T}_2\nonumber\\&&-3 \mathcal{T}_4+2 \mathcal{T}_5+2 \mathcal{T}_6-8 \mathcal{T}_8+2 \mathcal{V}_4-2
   \mathcal{V}_5-4 \mathcal{V}_6-r \mathcal{A}_1 x'_1-\mathcal{A}_1 x'_1-2 r \mathcal{A}_2 x'_1-2 \mathcal{A}_2 x'_1-2 r \mathcal{A}_3 x'_1-2 \mathcal{A}_3
   x'_1\nonumber\\&&+2 r \mathcal{A}_5 x'_1-4 r \mathcal{A}_6 x'_1+r \mathcal{P}_1 x'_1+2 r \mathcal{P}_2 x'_1-r \mathcal{S}_1 x'_1+2 r \mathcal{S}_2 x'_1+r \mathcal{T}_1
   x'_1+r \mathcal{T}_2 x'_1-3 r \mathcal{T}_4 x'_1+2 r \mathcal{T}_5 x'_1\nonumber\\&&+2 r \mathcal{T}_6 x'_1-8 r \mathcal{T}_8 x'_1+r \mathcal{V}_1 x'_1+\mathcal{V}_1
   x'_1-2 r \mathcal{V}_2 x'_1-2 \mathcal{V}_2 x'_1-2 r \mathcal{V}_3 x'_1-2 \mathcal{V}_3 x'_1+2 r \mathcal{V}_4 x'_1-2 r \mathcal{V}_5 x'_1-4 r
   \mathcal{V}_6 x'_1)\nonumber\\&&+\frac{2 r \phi _4 x'_3 M^3}{(r-1) (r+1)^2}(-\mathcal{P}_1 r^2+\mathcal{S}_1 r^2-\mathcal{T}_1 r^2-\mathcal{T}_2
   r^2+\mathcal{T}_4 r^2+\mathcal{P}_1 x'_1 r^2-\mathcal{S}_1 x'_1 r^2+\mathcal{T}_1 x'_1 r^2+\mathcal{T}_2 x'_1 r^2\nonumber\\&&-\mathcal{T}_4 x'_1 r^2+\mathcal{A}_1 r+2
   \mathcal{A}_3 r-\mathcal{P}_1 r+\mathcal{S}_1 r-\mathcal{T}_1 r+\mathcal{T}_2 r-\mathcal{T}_4 r-\mathcal{V}_1 r+2 \mathcal{V}_3 r+\mathcal{A}_1+2
   \mathcal{A}_3-\mathcal{P}_1\nonumber\\&&+\mathcal{S}_1-\mathcal{T}_1+\mathcal{T}_2-\mathcal{T}_4-\mathcal{V}_1+2 \mathcal{V}_3)-\frac{4
   r^2 \phi _{3 s} M^3}{(r-1) (r+1)^2}(-\mathcal{P}_1 x'_3 r^2-2 \mathcal{P}_2 x'_3 r^2\nonumber\\&&-\mathcal{S}_1 x'_3 r^2+2 \mathcal{S}_2 x'_3 r^2-\mathcal{T}_1 x'_3
   r^2+\mathcal{T}_2 x'_3 r^2+4 \mathcal{T}_3 x'_3 r^2+\mathcal{T}_4 x'_3 r^2-2 \mathcal{T}_5 x'_3 r^2+2 \mathcal{T}_6 x'_3 r^2-8 \mathcal{T}_7 x'_3\nonumber\\&&
   r^2+\mathcal{P}_1 x'_1 x'_3 r^2+2 \mathcal{P}_2 x'_1 x'_3 r^2+\mathcal{S}_1 x'_1 x'_3 r^2-2 \mathcal{S}_2 x'_1 x'_3 r^2+\mathcal{T}_1 x'_1 x'_3
   r^2-\mathcal{T}_2 x'_1 x'_3 r^2-4 \mathcal{T}_3 x'_1 x'_3 r^2\nonumber\\&&-\mathcal{T}_4 x'_1 x'_3 r^2+2 \mathcal{T}_5 x'_1 x'_3 r^2-2 \mathcal{T}_6 x'_1 x'_3 r^2+8
   \mathcal{T}_7 x'_1 x'_3 r^2+x_2 \mathcal{A}_1 x'_1 r+x_3 \mathcal{A}_1 x'_1 r+2 x_2 \mathcal{A}_4 x'_1 r+2 x_3 \mathcal{A}_4 x'_1 r\nonumber\\&&-2 x_2 \mathcal{A}_5
   x'_1 r-2 x_3 \mathcal{A}_5 x'_1 r-x_2 \mathcal{P}_1 x'_1 r-x_3 \mathcal{P}_1 x'_1 r-x_2 \mathcal{S}_1 x'_1 r-x_3 \mathcal{S}_1 x'_1 r-x_2 \mathcal{T}_1
   x'_1 r\nonumber\\&&-x_3 \mathcal{T}_1 x'_1 r+x_2 \mathcal{T}_2 x'_1 r+x_3 \mathcal{T}_2 x'_1 r+4 x_2 \mathcal{T}_3 x'_1 r+4 x_3 \mathcal{T}_3 x'_1 r-x_2 \mathcal{T}_4
   x'_1 r-x_3 \mathcal{T}_4 x'_1 r-4 x_2 \mathcal{T}_5 x'_1 r\nonumber\\&&-4 x_3 \mathcal{T}_5 x'_1 r-16 x_2 \mathcal{T}_7 x'_1 r-16 x_3 \mathcal{T}_7 x'_1 r+x_2
   \mathcal{V}_1 x'_1 r+x_3 \mathcal{V}_1 x'_1 r+2 x_2 \mathcal{V}_4 x'_1 r+2 x_3 \mathcal{V}_4 x'_1 r\nonumber\\&&-2 x_2 \mathcal{V}_5 x'_1 r-2 x_3 \mathcal{V}_5 x'_1
   r+\mathcal{A}_1 x'_3 r+2 \mathcal{A}_2 x'_3 r+4 \mathcal{A}_5 x'_3 r-\mathcal{P}_1 x'_3 r-2 \mathcal{P}_2 x'_3 r-\mathcal{S}_1 x'_3 r\nonumber\\&&+2 \mathcal{S}_2 x'_3
   r-\mathcal{T}_1 x'_3 r-\mathcal{T}_2 x'_3 r+4 \mathcal{T}_3 x'_3 r+3 \mathcal{T}_4 x'_3 r+2 \mathcal{T}_5 x'_3 r-2 \mathcal{T}_6 x'_3 r+8 \mathcal{T}_7
   x'_3 r\nonumber\\&&+\mathcal{V}_1 x'_3 r-2 \mathcal{V}_2 x'_3 r+4 \mathcal{V}_5 x'_3 r+2 x_2 \mathcal{A}_4+2 x_3 \mathcal{A}_4-2 x_2 \mathcal{A}_5-2 x_3
   \mathcal{A}_5-x_2 \mathcal{P}_1-x_3 \mathcal{P}_1-x_2 \mathcal{S}_1\nonumber\\&&-x_3 \mathcal{S}_1-x_2 \mathcal{T}_1-x_3 \mathcal{T}_1+x_2 \mathcal{T}_2+x_3
   \mathcal{T}_2+4 x_2 \mathcal{T}_3+4 x_3 \mathcal{T}_3-x_2 \mathcal{T}_4-x_3 \mathcal{T}_4-4 x_2 \mathcal{T}_5-4 x_3 \mathcal{T}_5\nonumber\\&&-16 x_2 \mathcal{T}_7-16
   x_3 \mathcal{T}_7+2 x_2 \mathcal{V}_4+2 x_3 \mathcal{V}_4-2 x_2 \mathcal{V}_5-2 x_3 \mathcal{V}_5+x_2 \mathcal{A}_1 x'_1+x_3 \mathcal{A}_1 x'_1+x_2
   \mathcal{V}_1 x'_1\nonumber\\&&+x_3 \mathcal{V}_1 x'_1+\mathcal{A}_1 x'_3+2 \mathcal{A}_2 x'_3+4 \mathcal{A}_5 x'_3-\mathcal{P}_1 x'_3-2 \mathcal{P}_2
   x'_3-\mathcal{S}_1 x'_3+2 \mathcal{S}_2 x'_3-\mathcal{T}_1 x'_3\nonumber\\&&-\mathcal{T}_2 x'_3+4 \mathcal{T}_3 x'_3+3 \mathcal{T}_4 x'_3+2 \mathcal{T}_5 x'_3-2
   \mathcal{T}_6 x'_3+8 \mathcal{T}_7 x'_3+\mathcal{V}_1 x'_3-2 \mathcal{V}_2 x'_3+4 \mathcal{V}_5 x'_3),
\end{eqnarray}

\begin{eqnarray}
H^{F_3}_a(x_i,x'_i)&=&\frac{2 r M^3\phi _2 (x'_1-1)}{(r-1) (r+1)^2}(x_2+x_3) (\mathcal{A}_1 r^2+2 \mathcal{A}_2 r^2+2 \mathcal{A}_4 r^2+2 \mathcal{A}_5 r^2+4 \mathcal{A}_6 r^2-2 \mathcal{T}_1 r^2+4
   \mathcal{T}_4 r^2-4 \mathcal{T}_5 r^2\nonumber\\&&+8 \mathcal{T}_8 r^2-\mathcal{V}_1 r^2+2 \mathcal{V}_2 r^2-2 \mathcal{V}_4 r^2-2 \mathcal{V}_5 r^2+4 \mathcal{V}_6
   r^2-2 \mathcal{A}_3 r+4 \mathcal{A}_5 r+\mathcal{P}_1 r+2 \mathcal{P}_2 r-\mathcal{S}_1 r\nonumber\\&&+2 \mathcal{S}_2 r-\mathcal{T}_1 r+\mathcal{T}_2 r+\mathcal{T}_4
   r-2 \mathcal{T}_5 r+2 \mathcal{T}_6 r-2 \mathcal{V}_3 r-4 \mathcal{V}_5 r-\mathcal{A}_1-2 \mathcal{A}_2-2 \mathcal{A}_3+\mathcal{V}_1-2 \mathcal{V}_2-2
   \mathcal{V}_3)\nonumber\\&&-\frac{2 r x'_3 M^3\phi _4 (x'_1-1)}{(r-1) (r+1)^2}
   (\mathcal{P}_1-\mathcal{S}_1+\mathcal{T}_1+\mathcal{T}_2-\mathcal{T}_4)\nonumber\\&&+\frac{4 \phi _{3
   s} M^3(x'_1-1)}{(r-1) (r+1)^2} (-x_2 \mathcal{A}_1 r^2-x_3 \mathcal{A}_1 r^2-2 x_2 \mathcal{A}_3 r^2-2 x_3 \mathcal{A}_3 r^2-2 x_2 \mathcal{A}_4 r^2\nonumber\\&&-2 x_3
   \mathcal{A}_4 r^2+2 x_2 \mathcal{A}_5 r^2+2 x_3 \mathcal{A}_5 r^2+2 x_2 \mathcal{T}_1 r^2+2 x_3 \mathcal{T}_1 r^2-8 x_2 \mathcal{T}_3 r^2-8 x_3
   \mathcal{T}_3 r^2+4 x_2 \mathcal{T}_5 r^2\nonumber\\&&+4 x_3 \mathcal{T}_5 r^2+16 x_2 \mathcal{T}_7 r^2+16 x_3 \mathcal{T}_7 r^2-x_2 \mathcal{V}_1 r^2-x_3
   \mathcal{V}_1 r^2+2 x_2 \mathcal{V}_3 r^2+2 x_3 \mathcal{V}_3 r^2-2 x_2 \mathcal{V}_4 r^2\nonumber\\&&-2 x_3 \mathcal{V}_4 r^2+2 x_2 \mathcal{V}_5 r^2+2 x_3
   \mathcal{V}_5 r^2+\mathcal{P}_1 x'_3 r^2+2 \mathcal{P}_2 x'_3 r^2+\mathcal{S}_1 x'_3 r^2-2 \mathcal{S}_2 x'_3 r^2+\mathcal{T}_1 x'_3 r^2\nonumber\\&&-\mathcal{T}_2
   x'_3 r^2-4 \mathcal{T}_3 x'_3 r^2-\mathcal{T}_4 x'_3 r^2+2 \mathcal{T}_5 x'_3 r^2-2 \mathcal{T}_6 x'_3 r^2+8 \mathcal{T}_7 x'_3 r^2-2 x_2 \mathcal{A}_3
   r\nonumber\\&&-2 x_3 \mathcal{A}_3 r-x_2 \mathcal{P}_1 r-x_3 \mathcal{P}_1 r-x_2 \mathcal{S}_1 r-x_3 \mathcal{S}_1 r+x_2 \mathcal{T}_1 r+x_3 \mathcal{T}_1 r+x_2
   \mathcal{T}_2 r+x_3 \mathcal{T}_2 r-4 x_2 \mathcal{T}_3 r\nonumber\\&&-4 x_3 \mathcal{T}_3 r-x_2 \mathcal{T}_4 r-x_3 \mathcal{T}_4 r+2 x_2 \mathcal{V}_3 r+2 x_3
   \mathcal{V}_3 r+x_2 \mathcal{A}_1+x_3 \mathcal{A}_1+x_2 \mathcal{V}_1+x_3 \mathcal{V}_1).
\end{eqnarray}

The corresponding formulas for the  axial-vector ones can be obtained by the following replacement:
\begin{eqnarray}
H^{G_i}_\xi&=&\pm H^{F_i}_\xi|_{r\to -r,\quad \phi _{3s}\to -\phi _{3s}},
\end{eqnarray}
where the plus and minus signs refer to  $i=1$ and $i=2,3$, respectively.

\begin{table}[H]
\footnotesize
\centering
	\caption{The expressions of  $\Omega_{\xi}$  in Eq.~(\ref{eq:FG}).}
	\label{tab:bb}
	\begin{tabular}[t]{lcc}
		\hline\hline
$\xi$  &$\Omega_{\xi}$\\ \hline
$a$  &$K_0(\sqrt{t_A}|\textbf{b}_2-\textbf{b}_3|)
       K_0(\sqrt{t_B}|\textbf{b}_3-\textbf{b}'_2+\textbf{b}'_3|)
       K_0(\sqrt{t_C}|\textbf{b}_2-\textbf{b}_3+\textbf{b}'_2-\textbf{b}'_3|)
       K_0(\sqrt{t_D}|\textbf{b}_3+\textbf{b}'_3|)$\\
$b$  &$K_0(\sqrt{t_A}|\textbf{b}'_2-\textbf{b}'_3|)
       K_0(\sqrt{t_B}|\textbf{b}_2|)
       K_0(\sqrt{t_C}|\textbf{b}_2-\textbf{b}_3+\textbf{b}'_2-\textbf{b}'_3|)
       K_0(\sqrt{t_D}|\textbf{b}_3+\textbf{b}'_3|)$\\	
 $c$   &$K_0(\sqrt{t_A}|\textbf{b}_3|) h_2(|\textbf{b}'_2|,|\textbf{b}_2+\textbf{b}'_2|,t_B,t_C,t_D) \delta^2(\textbf{b}_2-\textbf{b}_3+\textbf{b}'_2-\textbf{b}'_3)$\\
 $d$   &$K_0(\sqrt{t_A}|\textbf{b}_3-\textbf{b}_2-\textbf{b}'_2|)
         K_0(\sqrt{t_B}|\textbf{b}_2|)
         h_1(|\textbf{b}_2+\textbf{b}'_2|,t_C,t_D) \delta^2(\textbf{b}_2-\textbf{b}_3+\textbf{b}'_2-\textbf{b}'_3)$\\
 $e$  &$K_0(\sqrt{t_A}|\textbf{b}_2-\textbf{b}_3|)
       K_0(\sqrt{t_B}|\textbf{b}'_2|)
       K_0(\sqrt{t_C}|\textbf{b}_2-\textbf{b}_3+\textbf{b}'_2-\textbf{b}'_3|)
       K_0(\sqrt{t_D}|\textbf{b}_3+\textbf{b}'_3|)$\\
 $f$  &$K_0(\sqrt{t_A}|\textbf{b}'_2-\textbf{b}'_3|)
       K_0(\sqrt{t_B}|\textbf{b}_2-\textbf{b}_3-\textbf{b}'_3|)
       K_0(\sqrt{t_C}|\textbf{b}_2-\textbf{b}_3+\textbf{b}'_2-\textbf{b}'_3|)
       K_0(\sqrt{t_D}|\textbf{b}_3+\textbf{b}'_3|)$\\
 $g$   &$K_0(\sqrt{t_B}|\textbf{b}'_2|) h_2(|\textbf{b}'_3|,|\textbf{b}_2+\textbf{b}'_2|,t_A,t_C,t_D) \delta^2(\textbf{b}_2-\textbf{b}_3+\textbf{b}'_2-\textbf{b}'_3)$\\

 $h$  &$K_0(\sqrt{t_A}|\textbf{b}_2+\textbf{b}'_2-\textbf{b}'_3|)
       K_0(\sqrt{t_B}|\textbf{b}_2-\textbf{b}_3|)
       K_0(\sqrt{t_C}|\textbf{b}_2+\textbf{b}'_2|)
       K_0(\sqrt{t_D}|\textbf{b}_2-\textbf{b}_3+\textbf{b}'_2-\textbf{b}'_3|)$\\
$i$  &$K_0(\sqrt{t_A}|\textbf{b}_3|)
       K_0(\sqrt{t_B}|\textbf{b}'_2-\textbf{b}'_3|)
       K_0(\sqrt{t_C}|\textbf{b}_2+\textbf{b}'_2|)
       K_0(\sqrt{t_D}|\textbf{b}_2-\textbf{b}_3+\textbf{b}'_2-\textbf{b}'_3|)$\\	
 $j$   &$K_0(\sqrt{t_B}|\textbf{b}_2|) h_2(|\textbf{b}'_3|,|\textbf{b}_2+\textbf{b}'_2|,t_A,t_D,t_C) \delta^2(\textbf{b}_2-\textbf{b}_3+\textbf{b}'_2-\textbf{b}'_3)$\\
 $k$   &$K_0(\sqrt{t_A}|\textbf{b}_3|)
         K_0(\sqrt{t_B}|\textbf{b}'_2|)
         h_1(|\textbf{b}_2+\textbf{b}'_2|,t_C,t_D) \delta^2(\textbf{b}_2-\textbf{b}_3+\textbf{b}'_2-\textbf{b}'_3)$\\
 $l$  &$K_0(\sqrt{t_A}|\textbf{b}'_3|)
       K_0(\sqrt{t_B}|\textbf{b}_2-\textbf{b}_3|)
       K_0(\sqrt{t_C}|\textbf{b}_2+\textbf{b}'_2|)
       K_0(\sqrt{t_D}|\textbf{b}_2-\textbf{b}_3+\textbf{b}'_2-\textbf{b}'_3|)$\\
 $m$  &$K_0(\sqrt{t_A}|\textbf{b}_2-\textbf{b}_3+\textbf{b}'_2|)
       K_0(\sqrt{t_B}|\textbf{b}'_2 -\textbf{b}'_3|)
       K_0(\sqrt{t_C}|\textbf{b}_2+\textbf{b}'_2|)
       K_0(\sqrt{t_D}|\textbf{b}_2-\textbf{b}_3+\textbf{b}'_2-\textbf{b}'_3|)$\\
 $n$   &$K_0(\sqrt{t_A}|\textbf{b}'_3|) h_2(|\textbf{b}'_2|,|\textbf{b}_2+\textbf{b}'_2|,t_B,t_D,t_C) \delta^2(\textbf{b}_2-\textbf{b}_3+\textbf{b}'_2-\textbf{b}'_3)$\\
 \hline\hline
	\end{tabular}
\end{table}

\end{appendix}

\end{document}